\pdfoutput=1
\documentclass[acmsmall,nonacm]{acmart}

\usepackage{subcaption}
\usepackage{tabularx}
\usepackage{listings}

\usepackage{soul}
\usepackage[normalem]{ulem}
\AtBeginDocument{%
  }

\begin{document}

\title[Ethics Training Agents]{Ethics Training Agents: Facilitating Group-Based Ethics Education with Role-Playing and Discussion for Ethical Reflection and Exploration}

\author{Youngseok Seo}
\email{ysseo0910@kaist.ac.kr}
\orcid{0000-0002-1503-8939}
\affiliation{%
  \institution{KAIST}
  \city{Daejeon}
  \country{Republic of Korea}
}

\author{Sueun Jang}
\orcid{0009-0008-6029-5353}
\email{sueun.jang@kaist.ac.kr}
\affiliation{%
  \institution{KAIST}
  \city{Daejeon}
  \country{Republic of Korea}
}

\author{Hyesoo Park}
\orcid{0009-0008-8662-3378}
\email{hye@gatech.edu}
\affiliation{%
  \institution{Georgia Institute of Technology}
  \city{Atlanta, Georgia}
  \country{USA}
}
\author{Renz Samuel Gutierrez}
\orcid{0009-0002-4007-0727}
\email{renzgutierrez84@gmail.com}
\affiliation{%
  \institution{KAIST}
  \city{Daejeon}
  \country{Republic of Korea}
}
\author{Joseph Seering}
\orcid{0000-0001-7606-4711} %
\email{seering@kaist.ac.kr}
\affiliation{
  \institution{KAIST}
  \city{Daejeon}
  \country{Republic of Korea}
}

\author{Uichin Lee}
\orcid{0000-0002-1888-1569}
\email{uclee@kaist.edu}
\authornote{Corresponding author.}

\affiliation{%
 \institution{KAIST}
 \city{Daejeon}
 \country{Republic of Korea}}

\renewcommand{\shortauthors}{Seo et al.}

\begin{abstract}
Group-based ethics training for Science, Technology, Engineering and Mathematics (STEM) students is a complex challenge, requiring substantial resources and expertise. While activity-based teaching methods, such as role-playing and discussions, are commonly employed to simulate real-world scenarios, current practices are often manual and lack integration with effective online platforms for supporting group-based ethical discussions. In this work, we propose Ethics Training Agents, a group discussion system that leverages multiple LLM participants embodying distinct ethical orientations, along with a moderator agent, to enable structured human-AI group ethical discussions for collaborative reflection. We conduct a user study with 45 undergraduate STEM students to evaluate the learning outcomes and user experience. The results show that our system supports engagement, coordination, and perspective-taking in group discussions and has a positive influence on ethical sensitivity. We also discuss practical design strategies for integrating multiple LLM agents into multi-human group settings to facilitate ethics training for STEM students.
\end{abstract}



\authorsaddresses{}
\maketitle

\section{Introduction}
Engineering decisions have a broad societal impact, as technologies designed by engineers shape safety, privacy, equity, and sustainability in everyday life~\cite{Hess2018Systematic}. With the rapid growth of emerging fields such as AI and autonomous systems, engineers increasingly face novel dilemmas that extend beyond traditional technical problem-solving~\cite{Wang2024Farsight, Hanschke2024Data}. Yet, STEM curricula often prioritize technical mastery, leaving students underprepared to anticipate the societal consequences of their work. To address this gap, it is essential to cultivate ethical sensitivity and responsibility through early exposure to ethics~\cite{Friedman2019Value}. Such exposure helps learners recognize diverse stakeholders, anticipate unintended consequences, and lay the foundation for socially responsible engineering practice~\cite{Hess2018Systematic, Fiesler2020Teach}.

Existing educational materials~\cite{Brown2024Teaching} include both stand-alone courses (covering topics such as privacy, law, and fairness) and modular integrations within technical courses (e.g., ethics-informed assignments and debates). While broadening student exposure, these approaches often face challenges, as they are isolated as one-off assignments rather than being integrated into group activities. Furthermore, their effectiveness is highly dependent on the instructor's level of ethics expertise~\cite{Shapiro2021RolePlay, Brown2024Teaching}.

To bridge this gap, fictional role playing and discussion activities grounded in critical design and value sensitive design have been emphasized as promising pedagogical approaches, enabling integration into realistic, group-based activities~\cite{Bardzell2013Critical, Friedman2019Value}. Prior work has shown that role-playing can help surface ethical concerns through stakeholder perspective-taking, using speculative media (e.g., Black Mirror-based role plays~\cite{Klassen2022Run}) or game-based collaborative deliberation (Judgment Call~\cite{Ballard2019Judgment}). However, despite the well-documented benefits, it remains challenging to implement such activities at scale. Providing consistent learning experiences across multiple groups in classroom settings requires substantial instructor involvement and facilitation skills. Therefore, there is a need for digital tools that can scale role play-based ethics education~\cite{Pourghaznein2015, Jasemi2022}.

To address such challenges, digital tools for ethics training have also emerged. For example, PEaRCE~\cite{Castro2023Piloting} situates learners in fictional projects with structured feedback. Recent work proposes systems that employ generative AI to facilitate reflection~\cite{ElsayedAli2023Responsible} and foster multi-perspective understanding~\cite{Wang2024Farsight}. While valuable, these approaches remain largely unidirectional and individual-focused, providing limited support for collaborative ethics education. What is needed, therefore, are new group-based tools, embedding fictional role playing in collaborative ethical deliberation.

In this work, we extend Judgment Call~\cite{Ballard2019Judgment} into a multi-agent system\footnote{Throughout this paper, we use the term \emph{LLM agent} to refer to a system that employs LLMs as central controllers to construct autonomous agents to obtain human-like decision-making capabilities~\cite{Wang2024LLMAgentSurvey}. } where an LLM facilitator and multiple LLM participants with distinct ethical backgrounds guide discussions and enable ethical reasoning.
By incorporating an LLM facilitator into the system, we aim to reduce the facilitation and instruction overhead typically required to orchestrate role play activities at scale. Furthermore, by assigning consistent roles, personas, and value commitments to LLM agents, we can offer a stable and comparable set of perspectives across groups, supporting more consistent learning experiences across diverse teaching environments.
Unlike prior digital tools with limited group interaction~\cite{Castro2023Piloting, ElsayedAli2023Responsible, Wang2024Farsight}, our system situates ethics within a collaborative learning environment by supporting multi-human and multi-agent mixed role playing and discussions in group contexts. To understand the impact and design implications of such a system, we examine participants’ ethical sensitivity, facilitation effect of LLM agents, their assessment of LLM participants, and social dynamics in Human-AI Mixed ethical discussion. These considerations motivate the following research questions:
\begin{itemize}
\item RQ1. How do ethics discussions with LLM agents  and human peers affect participants’ ethical sensitivity?
\item RQ2. How does LLM agents' facilitation help learners engage in role-play ethics discussions?
\item RQ3. How do participants perceive the contributions of LLM agents compared to those of human participants?
\item RQ4. How do LLM agents shape social dynamics in multi-human, multi-agent ethical discussions?
\end{itemize}

\section{Related Work}

We review  traditions and approaches  of ethics education, highlight the benefits of fictional role playing, and provide an overview of recent tools for ethical technology design.
\subsection{Ethical Frameworks in Technology Design}
Ethical frameworks in technology design are often grounded in two main strands. First, philosophical theories such as consequentialism (judging actions by their outcomes), deontology (judging actions by adherence to rules or duties), and virtue ethics (judging actions by the character and virtues of the agent) provide foundational perspectives for evaluating the morality of technological actions~\cite{Zoshak2021Beyond, Kohno2023Ethical}.  More recently, scholars have highlighted the importance of diverse ethical traditions, including feminist care ethics~\cite{Henriques2025Feminist} and non-Western philosophies (e.g., Confucian moral philosophy~\cite{Lam2025Confucian}), to broaden the cultural and philosophical basis of ethics in computing~\cite{Zoshak2021Beyond}. However, these philosophical approaches can be abstract and difficult to translate into concrete design practices.

Therefore, as a second main strand, researchers in human--computer interaction have advanced value sensitive design (VSD), which operationalizes human values in moral and cultural contexts (e.g., autonomy, privacy, trust, well-being) by embedding them into the design process through principled methods such as stakeholder analysis, value scenarios, and multi-lifespan co-design~\cite{Friedman2019Value}. VSD thereby serves as a methodological bridge that reframes ethical frameworks as abstract ethics theories and translates human values into actionable design elements. Complementing VSD, scholars have also employed critical design, which surfaces hidden assumptions, value tensions, and ethical risks by creating provocative prototypes and speculative scenarios that invite reflection and debate~\cite{Dunne2013Speculative, Bardzell2013Critical}. VSD translates values into design requirements, while critical design provokes reflection on contested futures and hidden risks~\cite{Vilaza2022Scoping}.

Building on these research strands, corporations and multi-stakeholder organizations (e.g., Microsoft, Google, Partnership on AI) have articulated principle-based frameworks and exercises for computing and technology design that specify actionable guidelines such as fairness, transparency, accountability, inclusivity, safety, and privacy~\cite{Khan2022Ethics}. Among these exercises is Judgment Call, developed as a tabletop workshop game for industry teams to surface ethical concerns using VSD and design fiction methods~\cite{Ballard2019Judgment}. In this study,  we implement and evaluate a redesigned digital version of Judgment Call in multi-agent scenarios,  maintaining the VSD perspective that frames moral values as designable elements, thereby linking ethics education and technology design practice. We shift the target users of the game from experienced industry practitioners to students with little design experience by leveraging LLM agents to scaffold students' initial attempts to discuss ethical concepts in design.

\subsection{Ethics Education Practices in Technology Design}
Brown et al.~\cite{Brown2024Teaching} categorized existing ethics education efforts in technology design along two dimensions: (1) the strategy for ethics integration, either as dedicated courses or as modules embedded within existing courses (ranging from a single module, to multiple modules, or integration across the curriculum), and (2) the pedagogical approach, such as discussion, lecturing, or assignments. Dedicated courses on technology ethics typically cover topical areas including law, privacy, inequality, justice, human rights, fairness, and transparency~\cite{Fiesler2020Teach, Hess2018Systematic}.

Module-level integration has been widely adopted. For example, the Embedded EthiCS initiative incorporates active learning modules into computer science courses, enabling students to engage in ethical reasoning. In each course, instructors select a relevant ethical issue, introduce key concepts, and facilitate structured activities such as debates, discussions, and oral arguments grounded in philosophical principles~\cite{Grosz2019Embedded}. Similarly, recent work has explored various modular approaches: in introductory programming courses, students conduct ethics-informed coding assignments that prompt reflection on the social implications of their code~\cite{Jarzemsky2023Applies}; in algorithm design courses, students are asked to prioritize the safest path rather than the shortest path, aligning computational problem-solving with human values~\cite{Brown2022Shortest}.

However, Shapiro et al.~\cite{Shapiro2021RolePlay} highlighted that modular approaches face scalability challenges, both in terms of the number and breadth of courses across the curriculum. Also, prior work~\cite{pierrakos2019reimagining,Tran2024Student} has shown that current ethics education is compliance-oriented and loosely tied to ethical frameworks, teaching about ethical problems as if there is a single correct answer to be reached. As a result, students have few opportunities to consider the diversity of ethical concepts that may be relevant or to develop a coherent ethical perspective of their own. Furthermore, rather than focusing on realistic scenarios, many such educational modules rely on rare and extreme disaster cases that frame morality as an infrequent professional concern rather than an everyday disposition grounded in students’ disciplinary learning and design practice. This makes it difficult for students to connect ethical reasoning to what they learn in their courses and to apply it to realistic design situations.
 To address these limitations, subsequent studies~\cite{Shapiro2021RolePlay, schrier2017designing} have advocated topic-level role play as a promising approach to ethics education, as it can immerse students in more realistic and varied scenarios and encourage them to grapple with context-dependent trade-offs and stakeholder perspectives. Despite this promise, role play remains difficult to deliver at scale in practice compared to lecture based instruction. It requires substantial facilitation, time, and coordination overhead \cite{Pourghaznein2015, Jasemi2022}. Also, learning experiences can vary widely across groups depending on scenario setting, participant composition, and teaching environment, making it difficult to ensure consistent instruction and learning outcomes across groups. This practical gap motivates the need for tools that preserve the benefits of role play while reducing overhead and supporting consistent learning experiences.

Situated within these prior studies, our work adopts a module-level integration approach through a single module and employs discussion-based pedagogy with role playing for undergraduate students. To address the aforementioned challenges~\cite{Shapiro2021RolePlay, Tran2024Student}, we developed a multi-agent system based on Judgment Call~\cite{Ballard2019Judgment}, leveraging multiple LLM agents with distinct ethical and philosophical backgrounds to reduce facilitation overhead and support more consistent discussion experiences.

\subsection{Fictional Role Playing for Ethics Education}
A growing body of research has explored the use of fictional scenarios, role play, and game-based methods to engage students and practitioners in ethical reflection. Goldman’s Playbook for Ethical Technology Governance provides scenarios, decision trees, and guiding questions to help governments address ethical challenges in domains such as AI, climate, and public health, aligning governance with democratic values~\cite{Goldman2018Playbook}. Klassen and Fiesler~\cite{Klassen2022Run} similarly leveraged speculative fiction, using Black Mirror episodes to scaffold students’ ethical speculation about future harms and the societal consequences of present-day technologies.  Letters from the Future~\cite{Luria2022Letters} offers a structured five-step workflow (map, multiply, mediate, mount, re-map) to engage communities in imagining and experiencing future scenarios, creating feedback loops between speculative artifacts and ethical reflection. Role playing has also been incorporated  with VSD activities. Shapiro et al.~\cite{Shapiro2021RolePlay} demonstrated how role play can expose students to diverse stakeholder perspectives (technical experts, government officials, and the public).  In addition, game-based methods have been proposed to scale ethics integration and make ethical reasoning more engaging. DecidArch~\cite{DeBoer2019DecidArch, Alidoosti2023Ethics} is a software architecture game in which participants take on stakeholder roles to negotiate ethical issues using a deck of cards representing stakeholder interests, ethical values (e.g., dignity, privacy, accessibility), and potential system concerns. In industry contexts, Judgment Call~\cite{Ballard2019Judgment} supports product teams in surfacing ethical concerns through scenario-based role play that blends VSD and design fiction, prompting teams to deliberate trade-offs and document ethical decision-making.  More recently, So and Kim~\cite{So2024Dialogue} introduced AI Ethics Dilemmas, a card-based dialogue game where participants debate moral reasoning around AI ethics scenarios. Through role play, stakeholder framing, and structured discourse, the game enables participants to articulate diverse perspectives while competing to provide the most persuasive ethical arguments.

\subsection{Digital Tools for Ethics Education and Ethical Technology Design }
Digital ethics education and design tools offer structured interaction support, such as role playing and guided reflection, and systematic exploration of ethical issues with the help of database search or persona simulations. One feasible approach is to use role playing games where learners are asked to track and discuss their in-game decisions to enhance ethical awareness and reasoning~\cite{Schrier2014Designing, Schrier2015EPIC}. However, these game-based approaches are difficult to translate into concrete ethical design practices. To situate learners in realistic design contexts, in PEaRCE~\cite{Castro2023Piloting}, students act as employees in a fictional tech project, consult limited stakeholders, make decisions, and receive feedback via ethical coverage plots and real-world case links. AI LEGO~\cite{Wu2025AILEGO} extends this approach to AI design by enabling technical roles to draft lifecycle plans with interactive blocks and non-technical roles to review them through stage-specific checklists and LLM-based persona simulations, collaboratively identifying and mitigating potential harms in early design. In Responsible and Inclusive Cards~\cite{ElsayedAli2023Responsible}, technology practitioners engage in guided reflection through card-based questions on history, stakeholders, impacts, and actions across four phases (getting started, information gathering, getting specific, and call to action) to promote inclusive and harm-preventive design. LLM-based tools have also emerged to support the exploration of ethical issues. AHA!~\cite{Bucinca2023AHA} generates an ethics matrix of stakeholders and problematic AI behaviors, using LLMs and crowdsourcing to curate harm scenarios. In prompt-based prototyping, Farsight~\cite{Wang2024Farsight} helps detect potential risks by linking user prompts to past AI incident reports and auto-generating stakeholders, cases, and harms for critical review. The catalog of such AI-related incidents, such as the AI Incident Database (AIID), can also be used to raise awareness of risks and accountability~\cite{Feffer2023AIID}.

Taken together, these digital tools support ethics education and ethical technology design through individual reflection or the use of artifacts (e.g., checklists, matrices, and incident reports). While effective for helping individuals surface issues, they provide limited support for interactive, multi-party ethical deliberation. Such deliberation is crucial in ethics education because ethical design decisions are often made within teams, where participants work through ethical tensions together in real time~\cite{Tran2024Student}. Prior work on team-based ethical tools has largely examined either fully human teams~\cite{Alidoosti2023Ethics} or a single human orchestrating multiple LLM agents~\cite{Wu2025AILEGO}; to the best of our knowledge, ethical discussions that bring together multiple humans and multiple LLM agents remain underexplored.
Building on Judgment Call~\cite{Ballard2019Judgment}, our system introduces a collaborative, multi-human and multi-agent setting for role play-based ethics education. Moving beyond individual reflection or artifact-centered, limited group interaction, the system places students in a team-based deliberation process where multiple humans engage with multiple LLM participants to exchange counterarguments, negotiate perspectives, and converge on a group decision. This mixed setting closely aligns with the team-based nature of ethical design decision-making in practice.
\section{System Design}
We illustrate the structure of discussion, describe the design of the facilitator and participant agents, and present the interface design.
We used the GPT-4o model to implement the LLM-based agents, as it was a widely used, high-performing model for human-facing conversational and generative tasks at the time of the study~\cite{OpenAI2024GPT4o}.

\subsection{Authors' Positionality}
To provide transparency regarding the design and evaluation of this system, we briefly describe our research team's positionality in ethics education. Our team includes both faculty and graduate students in human--computer interaction, social computing, and ubiquitous computing, and includes members with more than 15 years of combined experience in incorporating ethical education into computer science courses, primarily at the undergraduate level in HCI courses. We acknowledge that the cultural context in which this study was conducted (East Asia) may have shaped our design choices and results.

\subsection{Discussion Structure}
\begin{figure}[t]
    \centering
    \includegraphics[width=\linewidth]{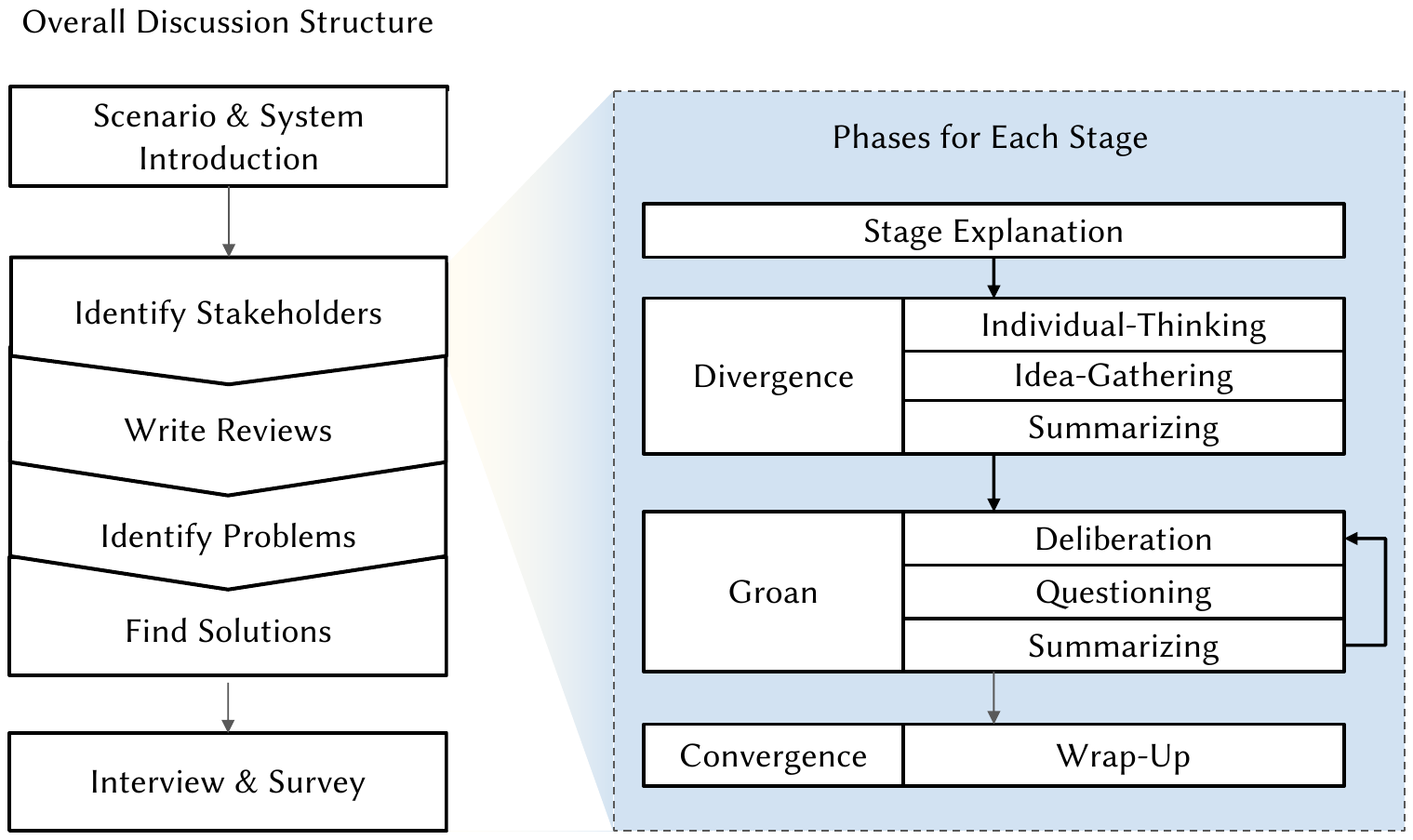}
    \caption{The study procedure, including the overall discussion structure (left) and the general discussion flow for each stage (right)}
    \Description{Flow diagram with two panels. The left panel shows the study timeline: beginning with Scenario & System Introduction, leading into four discussion stages (Identify Stakeholders, Write Reviews, Identify Problems, and Find Solutions), followed by Interview & Survey. The right panel shows the repeated structure of a discussion stage: starting with Stage Explanation, followed by Divergence (Individual-Thinking, Idea-Gathering, Summarizing), then Groan, where Deliberation and Questioning alternate in a loop before Summarizing, and ends with Convergence, which is the Wrap-Up.}
    \label{fig:Discussion Structure}
\end{figure}

\subsubsection{Judgment Call as a Foundation Structure}
We decided to ground our human-AI group ethical discussion system in an existing ethical computing and technology design activity that could serve as a structural foundation. Specifically, the tool needed to satisfy the following four criteria to facilitate ethical discussions: (1) it should be a team-based, discussion-oriented activity; (2) it should reflect the principles of VSD through role playing; (3) it should provide structured, well-specified objectives that an LLM can follow reliably; and (4) it should be completable within the time constraints of a single university class session, so that it could be adopted at scale within existing curricula. Based on these criteria, we selected Judgment Call~\cite{Ballard2019Judgment} as  a foundational structure. Guided by the fourth criterion, we constrained the total discussion activity to approximately 75 minutes including the introduction and the explanation of discussion stages, matching the typical length of a class period at our institution.

Judgment Call~\cite{Ballard2019Judgment} is comprised of four different stages %
as described below (Fig. \ref{fig:Discussion Structure}). Time allocated for each stage follows Judgment Call's original design. The introduction session and the explanation of the discussion stages were each allocated 5 minutes.
\begin{enumerate}
    \item \textbf{Identify Stakeholders  (10-15 min)}: Participants discuss and identify the stakeholders likely to be affected by the technology in question, including direct, indirect, and excluded stakeholders.
    \item \textbf{Write Reviews  (10-15 min)}: Each participant is randomly assigned a role comprising a stakeholder (identified from the previous stage), an ethical value, and a stance following Judgment Call's design. Based on this assigned role, they write 1--2 reviews regarding the technology.
    \item \textbf{Identify Problems  (10-15 min)}: Participants discuss recurring concerns and ethical issues that have surfaced through the review writing stage. At this stage, participants identify the most pressing problems, such as bias, exclusion, safety risks, or privacy concerns.
    \item \textbf{Find Solutions  (10-15 min)}: Participants propose potential solutions to the identified problems. These may include technical design modifications, policy interventions, or social strategies to mitigate harm and strengthen the benefits of the technology.
\end{enumerate}

\subsubsection{Phases for Each Stage}

To facilitate a structured discussion, each stage of our system  consisted of three phases: \textbf{Idea-gathering} for divergence, \textbf{Deliberation and Questioning} for groan, and \textbf{Wrap-up} for convergence. This aligns with the model of participatory decision-making, which comprises phases of divergence (generating diverse ideas), groan (struggling with differences), and convergence (narrowing down to shared decisions)~\cite{baker1999facilitator}. The specific procedures for each phase were adapted from Kaner et al.~\cite{kaner2014facilitator}.
These phases are applied to all discussion stages except for the \textit{Write Reviews} stage, where participants individually write reviews based on their assigned roles.

\begin{enumerate}
    \item \textbf{Idea-gathering}: This phase is open to all participants and lasts for 90 seconds at the beginning of each stage. It aims to provide participants with opportunities to freely present casual ideas that come to mind to enrich the discussion.
    \item \textbf{Deliberation}: Participants who want to add further opinions are given the floor to speak one by one. This phase is designed to encourage more deliberate explanations than the \textbf{Idea-gathering} phase. Therefore, there is no time constraint for this phase.
    \item \textbf{Questioning}: Participants who want to ask questions about others’ opinions are given the floor to speak. In response, one of the participants is selected to respond and is given the floor with unlimited time. The \textbf{Deliberation} and \textbf{Questioning} phases repeat alternately until the discussion time for the stage ends.
    \item \textbf{Wrap-up}: Before proceeding to the next stage, the key ideas are selected based on the criteria of Judgment Call~\cite{Ballard2019Judgment} (e.g., most impactful, recurring in discussion) and presented to the participants. These selected ideas are used in the next stages.
\end{enumerate}

\subsection{LLM Agents: The Facilitator}
In the original Judgment Call game~\cite{Ballard2019Judgment}, one of the participants acts as a facilitator to moderate the discussion and encourage engagement. Adapting this to a scalable multi-agent system, we designed an LLM agent which has three main roles as a facilitator: \textbf{Flow Control}, \textbf{Summarization}, and \textbf{Idea Selection}.

\subsubsection{Flow Control}
The facilitator should manage the discussion flow and phase transitions to prevent dominant voices or premature convergence of ideas~\cite{kaner2014facilitator}. Specifically, the facilitator manages speaking turns through \textbf{Stacking} and \textbf{Designating an answerer}, as well as \textbf{Time management} throughout the discussion.

\begin{itemize}
    \item \textbf{Stacking}: To ensure equal opportunities to speak, we adopted stacking~\cite{kaner2014facilitator}, which allows participants wishing to speak to raise their hands in advance. When the \textit{Deliberation} or \textit{Questioning} phase begins, the facilitator invites the participants to speak by asking them to press the raise hand button within 15 seconds. The facilitator then randomly assigns speaking order to those who raised their hands and informs the participants of this order, after which they express their opinions or questions accordingly. In addition, the facilitator encourages participation when no one raises their hand.
    \item \textbf{Designating an answerer}: During the \textit{Questioning} phase, the facilitator determines who answers each question. If a participant asks a question to a specific person, the designated person is given the floor to respond. If the question concerns a particular idea, the person who proposed the original idea or added a related comment is given the floor. For questions not directed at a specific participant or idea, the facilitator either answers the question itself or randomly selects one participant to provide a response. To simplify the process, the facilitator agent is designed to elicit one answer per question.
    \item \textbf{Time management}: The facilitator agent is designed to manage time throughout the discussion. For example, in the \textit{Idea-gathering} phase, after the allotted 90 seconds, no further inputs are accepted, and the facilitator announces that the phase has ended. The facilitator also checks how much time has elapsed within each stage. If the elapsed time exceeds 9 minutes, the facilitator directs the discussion to the \textit{Wrap-up} phase; otherwise, the facilitator alternates between the \textit{Deliberation} and \textit{Questioning} phases.

\end{itemize}

\subsubsection{Summarization}

To help participants recall details from the previous discussion, the facilitator agent is designed to summarize ideas and present them to the participants at the end of each phase.
When the system was prompted to summarize the entire discussion in a single query, we initially encountered problems, such as omitting ideas raised during the discussion or generating non-existent ideas, causing hallucination. In addition, each time the discussion content was summarized all at once, the summary changed drastically, making it difficult for participants to follow the discussion flow.

Prior work on LLM-based qualitative analysis systems addressed these issues by introducing a code analysis agent that examined data one piece at a time, checked for redundancy, and added a new code only when no overlap was found~\cite{katz2024thematic}. Inspired by this, we implemented the system to repeatedly examine the summary set, check whether a similar idea already existed, and add a new entry only if no overlap was found.

Furthermore, during the discussion, participants often added comments or clarified their intentions regarding existing ideas, which required modifications to the summary instead of new additions. To ensure that the summary reflects the discussion update, after generating a sufficiently large set of summaries, the system runs a validation step in which it repeatedly examines whether any ideas need to be revised.

Finally, running the summarization module for every participant's response would have been too time-consuming. Therefore, we performed summarization in batches at the end of each phase.

\subsubsection{Idea Selection}
At the end of each discussion stage, the brainstormed ideas usually exceeded the number that participants could discuss sufficiently, making it difficult to address all the ideas within a limited time.
Therefore, the facilitator agent is designed to select only a subset of ideas based on criteria adapted from the four questions in \emph{Judgment Call}: i.e., the most impactful, the most problematic, those with existing potential harm, and those recurring in discussion~\cite{Ballard2019Judgment}.

During the \textit{Identify Stakeholders} stage, the facilitator selects up to 10 stakeholders for divergence. For the \textit{Write Reviews} stage, no limit is placed on the number of reviews. In the \textit{Identify Problems} stage, the facilitator selects up to 8 problems.
In the \textit{Find Solutions} stage, the facilitator selects up to 8 important solutions. Instead of converging on a single best solution as in the original game~\cite{Ballard2019Judgment}, multiple solutions are provided as the final outcomes.

After the idea selection, the facilitator summarizes the chosen ideas and announces them to the participants before guiding them to the next stage.

\subsection{LLM Agents: The Participants}
All participant agents are designed to engage in every discussion phase: Ideation and Deliberation, Question and Answer, and Review. For each agent, we explicitly specified the values it should foreground during deliberation and instructed it to consistently reflect those values in its contributions, following VSD~\cite{Friedman2019Value}. In addition, to prevent the inherent unpredictability of LLM outputs from disrupting the discussion, we constrained agents to the objective of the current stage and explicitly prohibited responses that deviated from that goal (e.g., introducing solutions during the Identify Stakeholders stage).

\subsubsection{Ideation and Deliberation}
In the \textit{Idea-gathering and Deliberation} stages, we prompted the LLM agents to generate ideas in accordance with the discussion flow. Each agent was provided with information regarding both the overall discussion structure and the current stage. In addition, we instructed the agents to propose ideas that differed from those already presented to enrich the discussion. Furthermore, we prompted them to generate phase-specific responses: for example, during the \textit{Idea-gathering} phase, we prompted them to generate brief ideas consisting of one to three words, whereas in the \textit{Deliberation} phase, we prompted them to provide one idea accompanied by an explanation to encourage deeper and more elaborate suggestions.

\subsubsection{Question and Answer}
When generating questions in a single step, agents often produced errors, such as asking about advanced domain-specific knowledge, posing questions irrelevant to the current stage, or even asking questions to themselves. To address this, we divided the prompting into two steps: select an idea the persona might naturally be curious about, and then generate a question about that idea. Specifically, we provided the full discussion history of the current stage and prompted the agents to select one idea that might be curious or interesting to them and to explain the reason for their curiosity. Then, based on the selected idea, we prompted the agent to formulate a question in simple, friendly language aligned with the current discussion stage.
For prompts designed to respond to other participants’ opinions, the types and content of questions could not be predicted in advance; thus, we did not explicitly define the response format beyond basic discussion guidelines (e.g., the agent's persona, current stage, and the specific question).

\subsubsection{Review}
For review writing, we prompted the LLM agents to follow the guidelines from Judgment Call~\cite{Ballard2019Judgment}.
When composing a review, each participant is instructed to first imagine potential features of the product, then imagine user experiences based on those features, and finally write reviews grounded in those imagined experiences. Following this, the LLM agents proceeded in three steps: generating hypothetical features, generating hypothetical experiences, and then composing a review. More specifically, each agent was prompted to generate features and experiences that their assigned persona was likely to imagine and reflect the persona's information in the review writing.

\subsubsection{\texorpdfstring{Iterative Design}{Iterative Design}}
\label{sec:iterativedesign}

Before the main study, we conducted iterative pilot testing biweekly. For the first three sessions, one participant used the system at a time to evaluate the overall UI/UX, to debug system errors, and to refine agent prompts. For example, we added system messages and discussion topic explanations throughout this process as pilot testing revealed points of confusion. In the following sessions, six participants engaged in full group discussions to test the discussion structure itself. We conducted a total of five of these full testing sessions until participants reported that they were satisfied with the discussion structure and quality of LLM agent responses.

During pilot sessions, we observed the facilitator and participant agents occasionally referencing ideas or exchanges that had not actually occurred in the discussion, which prior work shows is more likely at higher temperatures~\cite{Zouaghi2025Temperature}. We therefore set the temperature to 0 for all agents, aiming to minimize user confusion. This also had the benefit of reducing variability across experimental sessions, which was important given our between-group comparisons.

We also calibrated the agents’ speaking speed, utterance length, and participation rate in order to achieve a relatively natural-feeling discussion flow. In our system, human users were made fully aware that the LLM agents were in fact AI agents and not human participants, but pilot testers still expressed a preference for LLM agents to communicate in a manner relatively similar to human patterns of conversation. This took several forms.

First, since LLM agents can generate utterances much faster than humans can type,  pilot participants expressed discomfort with sudden large volumes of information posted by the LLM agents. To address this, we introduced a pre- and post-delay around each agent’s utterance. The pre- and post-delays were initially calibrated to approximate human reading and typing speeds (reading: 150 words per minute; typing: 60 words per minute in Korean~\cite{Byun2018Performative, Kim2016Korean}). However, pilot participants reported that this setting felt too slow. Accordingly, the typing delay was adjusted to match the reading speed, calculated by dividing the word count by a reading speed of 150 words per minute.

Second, pilot participants reported cognitive burden when reading long messages from the LLM. Reflecting this feedback, we set an upper limit of 150 characters per message to prevent LLM agents' responses from becoming too lengthy. Moreover, we split the agents’ messages into shorter speech bubbles using periods as separators to guide participants' reading and facilitate understanding.

Third, we tested the appropriate participation rate of the LLM agents to ensure that participants were neither overwhelmed by excessive AI participation nor deprived of learning opportunities due to minimal AI involvement. To determine the appropriate participation rate in stacking (i.e., how actively the LLM agents should volunteer to speak), we tested participation rates from 0.2 to 0.8 during pilot sessions. Participants reported that the agents' activeness was appropriate regardless of the chosen rate;  %
therefore, we set the rate at 0.5 for the main user study.

\subsubsection{Persona Setting}
To provide students with a comprehensive and diverse set of ethical perspectives, we selected three complementary frameworks---\emph{care ethics}, \emph{deontological ethics}, and \emph{pragmatic ethics}---as the basis for our multi-agent system. This configuration is designed to expose students to distinct yet interconnected moral approaches.
In pilot sessions, we observed that instructing the LLM with naive ethical frameworks (e.g., “adopt a deontological perspective”) often produced responses that were underspecified, leading to responses with complex or vague language. To make each perspective actionable, we operationalized it as a persona with concrete behavioral guidance: we specified demographic cues (age, gender, major), writing style, a brief ethical background, and a set of value prioritizations following VSD guidelines. This helped the agents express their positions in distinctive and specific ways.

\begin{enumerate}
    \item \textbf{Care Ethics (Yeon-Su Song)}: This LLM agent is characterized as a student in social welfare who has experience in community volunteering. She emphasizes inclusivity and the importance of preventing anyone from being marginalized.
    \item \textbf{Deontological Ethics (Hoon Park)}: This LLM agent is characterized as a philosophy student with a background in political science. He emphasizes duties and principles, often stressing the need to consider worst-case scenarios in ethical decision-making.
    \item \textbf{Pragmatic Ethics (Mu-Young Lim)}: This LLM agent is characterized as an engineering student with a minor in media art and technology. He emphasizes practicality and feasibility, often highlighting efficiency and realistic solutions in ethical discussions.
\end{enumerate}
To reliably construct personas that represented a designated perspective, we followed Hwang et al.  \cite{hwang2023aligning} and composed each persona prompt into demographics, ideology, and opinions. We constructed ideology and opinions based on the designated ethical perspectives described above. For demographics, we specified the agents as undergraduate students so that participants would regard them not as experts whose opinions should be deferred to, but as peers with whom they can raise questions, challenge, and learn together. Each agent's age was randomly determined from a range of 19 to 26.

To validate the personas, the research team reached a consensus on a list of three general ideas that each persona was expected to contribute at every stage. We simulated discussions with the three agents, with the participation rate temporarily set to 1.0 (instead of the 0.5 noted above), in order to require them to participate in every section for testing purposes. Discussions proceeded with one idea gathering phase and three deliberation-questioning rounds. We examined whether all pre-specified ideas were raised. Because every stage except the stakeholder stage takes the preceding stage's output as input, we supplied the three ideas generated by researchers as input for each stage.

Throughout iterative testing, we repeatedly asked participants whether each persona was perceived as distinctive and representative of the designated perspective. We proceeded to the main experiment only after pilot participants reported that agents represented the designated perspective consistently. Note that, throughout this process, we consistently audited agent output logs for safety issues but did not encounter any, possibly due to the relatively constrained nature of agent participation, and no inappropriate agent behaviors were documented during the subsequent user studies. However, we acknowledge that it is not possible to guarantee with certainty that agent behaviors will always fall within appropriate boundaries. %
To mitigate this residual risk, a trained facilitator was available throughout each study session to intervene if necessary, and participants were reminded that they could pause or withdraw from the study at any time without penalty.

\subsection{User Interface}
\begin{figure*}[t]
    \centering
    \includegraphics[width=\linewidth]{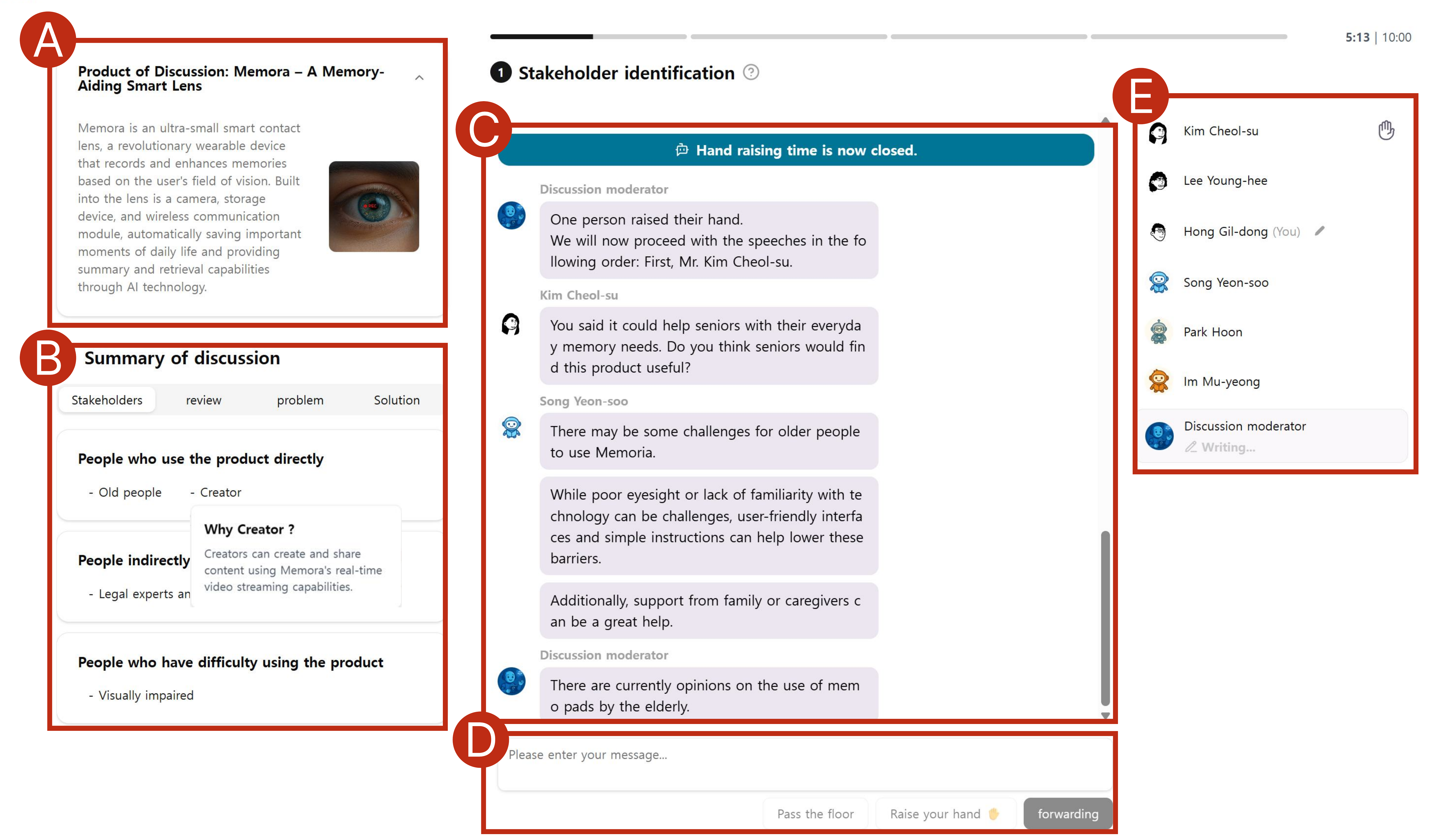}
    \caption{The user interface of the multi-human, multi-agent ethical discussion system. Participants can see the technology and scenario under discussion (A). The summary board (B) presents facilitator-generated summaries. The discussion content (C) is displayed in the center in chronological order. The message input area (D) provides a text box and interaction buttons. The status panel (E) shows who holds the speaking turn.}
    \Description{Screenshot of the ETA web interface with labeled regions A-E. (A) The left panel shows the technology and scenario prompt under discussion, along with a small image thumbnail. (B) A "Summary of Discussion" board lists facilitator-generated bullet summaries grouped by category. (C) The center panel shows the chronological chat for the current stage, with a banner indicating whether hand-raising is open or closed. (D) The bottom input area contains a text field and buttons for sending messages and raising a hand. (E) The right sidebar lists participants (humans, LLM personas, and a moderator) and indicates who currently has the speaking turn or has raised a hand.}
    \label{fig:userInterface}
\end{figure*}
We implemented a web application that allows users to communicate with other humans and LLM participants for ethical discussion. The technology and scenario under discussion are displayed in the top-left corner (Fig. \ref{fig:userInterface} A), allowing participants to keep track of the given context. Summaries generated by the facilitator agent are shown in the summary board (Fig. \ref{fig:userInterface} B), where each idea is represented by a short keyword and category; hovering over an item reveals a more detailed description.
Participants can also revisit the summaries from previous discussion stages by selecting the stage name. The center of the screen presents the discussion content (Fig. \ref{fig:userInterface} C), displaying the conversation in chronological order.
Participants can write their opinions in the message input area (Fig. \ref{fig:userInterface} D). This area includes interaction buttons as well as a text box: a button to send a message, a button to pass the turn after finishing an utterance, and a raise hand button to request the floor when the facilitator invites participation through stacking. %
Lastly, the participant status panel (Fig. \ref{fig:userInterface} E) shows who is currently typing, who holds the speaking turn, and who has raised a hand, making the overall state of participation visible to everyone.

\section{Implementation}
We implemented a system with three modules: a web client to render the user interface, a WebSocket server that synchronizes discussion state between users and agents, and an agent orchestration module that manages agents' behavior following the discussion structure. We implemented each stage and phase as an explicit state node using the LangGraph frameworks. Detailed explanation regarding implementation can be found at Appendix~\ref{sec:program_structure}.
\section{User Study Methods}

\subsection{Participants}
A total of 45 participants were recruited through a university online community.
The participants were assigned to 15 groups, each consisting of three individuals. %
All participants were undergraduate students majoring in STEM fields. Freshmen were excluded to ensure that participants had sufficient knowledge and experience in technology design through their major coursework.
Participants were 42\% female and 58\% male (Age: $M$ = 22.1, $SD$ = 1.7).
This study was reviewed and approved by the authors’ institutional review board (IRB). All participants provided informed consent prior to participation and could withdraw at any time without penalty. %

\subsection{Study Procedures}
The experiment was conducted offline in a university classroom, where participants interacted with the group discussion system for approximately two hours. Each participant was paid KRW 40,000  (approximately USD 28.8) as compensation. The study procedure was structured as follows:

\begin{itemize}
    \item \textbf{Pre-Survey (10 min):} Participants completed a short pre-survey to gather demographic information and baseline data.
    \item \textbf{Introductory Session (10 min):} Participants were briefed on the study objective, procedure, and the specific technology and scenario for discussion. They were also introduced to the personas of the LLM agents\footnote{With participants, we used the term ``AI'' rather than ``LLM agent'' throughout the study materials, introductory session, and interviews, as it was more accessible to a general, non-technical audience. Direct participant quotes in this paper therefore retain participants' own wording (e.g., ``AI''), while our own analysis and reporting use the term \emph{LLM agent} as defined in Section 1.} that would participate in the discussion, as well as to the functionalities of the discussion system.
    \item \textbf{Experiment (60 min):} The main experiment was organized into multiple stages. For each stage, participants were first provided with the stage objective (2 min) and reflected individually (2 min). They then engaged in a 10--15 minute ethical discussion with both human and LLM agents on the system. This cycle was repeated across the discussion stages.
    \item \textbf{Post-Survey (10 min):} Following the experiment, participants completed a post-survey evaluating their experience with the system and the discussion.
    \item \textbf{Interview (30 min):} Participants participated in semi-structured focus group interviews to provide qualitative feedback.
\end{itemize}

\subsection{Scenario}
We provided a specific technology and scenario as the topic for discussion, offering a consistent foundation for the randomly matched groups. Following prior work that employed Black Mirror episodes as design fiction scenarios for ethics education~\cite{Klassen2022Run}, we selected an episode from Black Mirror that featured a smart lens device with a miniature camera for memory augmentation. Participants engaged in group discussion tasks as if they were the engineers designing this technology. The detailed scenario prompt is described in Appendix \ref{sec:scenario}.

\subsection{Data Collection and Analysis}
To examine the educational impact of group ethical discussions involving LLM agents (RQ1), we collected ethical sensitivity scores before and after the discussion. To understand participants' perceptions and evaluations of human and LLM agents (RQ3), we  collected peer ratings after the discussion. We also collected discussion logs to examine interaction between participants and agents (RQ4).  Furthermore, we gathered qualitative data from focus group interviews to understand the participants' nuanced experiences (RQ1--RQ4).

\subsubsection{Ethical Sensitivity in Technology Design}
Drawing on prior work on moral development~\cite{jagger2011ethical}, we conceptualize ethical sensitivity in technology design as \emph{ethical issue awareness} that captures learners’ sensitivity to the presence of ethically salient features in technology design situations, and \emph{ethical consequence reasoning} that reflects more deliberate reasoning about potential consequences and alternative courses of action following such recognition.
Thus, we designed a measurement instrument conceptually grounded in the ethical sensitivity scale~\cite{Tirri2011ESS} to examine how our group discussion system improved ethical sensitivity (RQ1), as detailed in Table~\ref{tab:Ethical sensitivity} in the Appendix. For example, our instrument includes items such as “I am aware that product design can unintentionally disadvantage certain groups” to capture ethical issue awareness, and “I can identify possible downstream consequences of technology decisions” to capture ethical consequence reasoning.

\subsubsection{Peer Rating}
To evaluate the relative contributions and perception of human and LLM agents (RQ3), we collected peer ratings across three dimensions: contribution, diversity, and influence. To assess participant contribution, we adapted the CMU Eberly Center’s peer evaluation framework for group work~\cite{cmu_groupwork}. These ratings measured the extent to which each participant contributed to the overall discussion process. Furthermore, to assess whether each LLM agent effectively embodied its assigned persona and provided multiple perspectives, we adapted the rubrics for discussion~\cite{simmons2018sample}. Based on these rubrics, participants were evaluated on diversity (the extent to which they offered distinct perspectives) and influence (the extent to which they shaped participants’ thinking).
Participants were instructed to rate the other two human participants and the three LLM agents on a four-point Likert scale, individually and privately. The detailed questionnaire items are presented in Table \ref{tab:Peer Rating} of Appendix.

\subsubsection{Data Analysis}
We employed both qualitative and quantitative data analysis methods.
For the interview data, we conducted a reflexive thematic analysis~\cite{braun2006using}. We iteratively identified, refined, and interpreted themes that emerged from participants’ experiences and reflections during the focus group interviews. For the survey data, we used two main statistical analyses. Because both measures are composite scores derived from multiple ordinal Likert scale items, for each measure we examined whether they satisfied a normality test before selecting a parametric or non-parametric test~\cite{harpe2015analyze}. First, to examine changes in ethical sensitivity, we conducted paired-sample t-tests. We verified the normality by conducting Shapiro--Wilk tests on the pre--post difference scores for each sub-construct. Both ethical issue awareness ($W=0.972$, $p=.382$) and ethical consequence reasoning ($W=0.979$, $p=.635$) passed a normality test. As a robustness check, we additionally conducted Wilcoxon signed-rank tests on both sub-constructs, which corroborated the $t$-test results (see Section \ref{sec:RQ1_Result}).
Second, for peer rating data,  we similarly conducted Shapiro--Wilk tests.  The peer rating measures failed the normality tests; therefore, we used the nonparametric Kruskal--Wallis test to examine group differences. For all quantitative metrics adapted for our study, we evaluated internal consistency using Cronbach’s $\alpha$. All scales showed acceptable internal consistency ($\alpha$ > 0.70)~\cite{tavakol2011making}.
Lastly, for the interaction logs, we tested whether participants interacted more often with LLM agents or human participants via the number of questions they asked. Since an asker cannot address themselves, each participant could address three agents but only two human peers, giving a baseline of a 3:2 ratio of questions. Because exchanges are nested within groups, we treated the group as the unit of analysis and tested the group-level proportion against 3 : 2 with a Wilcoxon signed-rank test.

\section{Results}

Results are reported in relation to our four research questions (RQ1--RQ4). We first examine changes in participants’ ethical sensitivity (RQ1), followed by the roles of our system's design components in facilitating ethical discussions (RQ2), participants’ assessments of LLM agents' contributions (RQ3), and finally the social dynamics in human-AI mixed ethical discussions (RQ4). For clarity, direct quotes from participants are labeled as G[1–15]P[1–3], with G representing the group number and P representing the participant’s index within the group.

\subsection{RQ1. Changes in Ethical Sensitivity in Technology Design}
\label{sec:RQ1_Result}

\begin{figure}[t]
    \centering
    \begin{subfigure}[t]{0.45\linewidth}
        \centering
        \includegraphics[width=\linewidth]{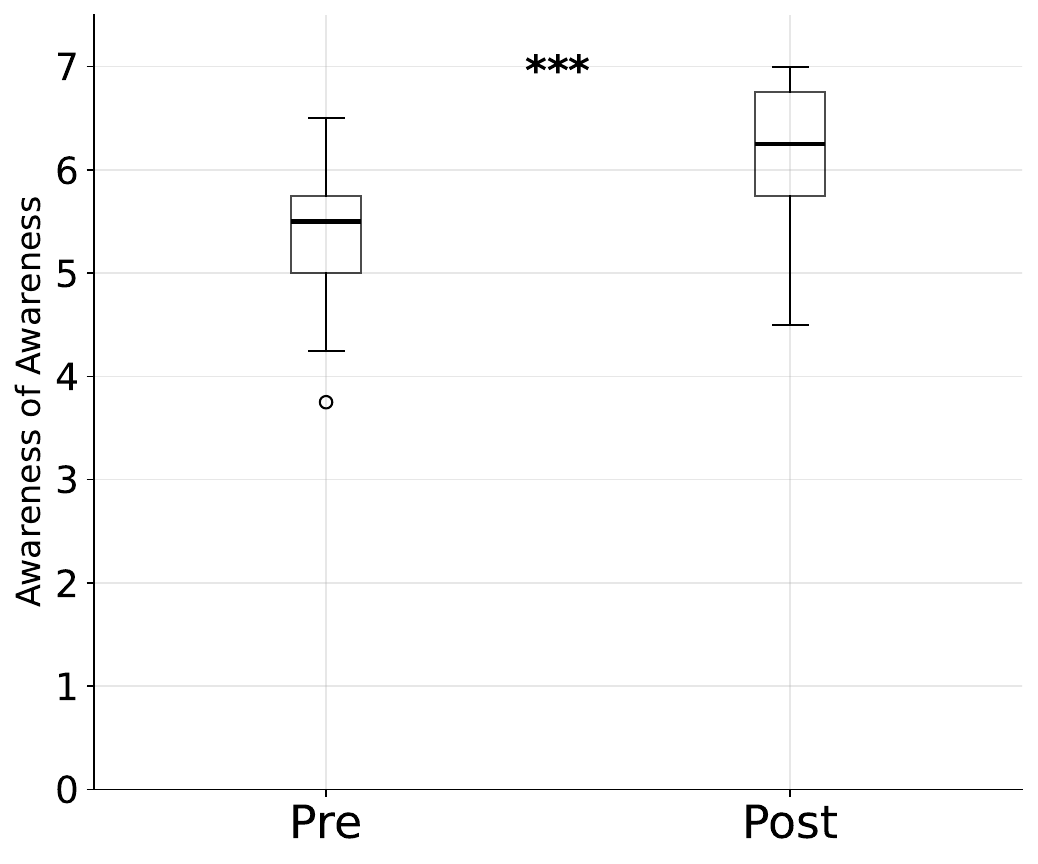}
        \Description{Box plot of ethical issue awareness scores on the y axis and time points (pre/post) on the x axis}
        \caption{Ethical issue awareness}
    \end{subfigure}
    \hspace{4mm}
    \begin{subfigure}[t]{0.45\linewidth}
        \centering
        \includegraphics[width=\linewidth]{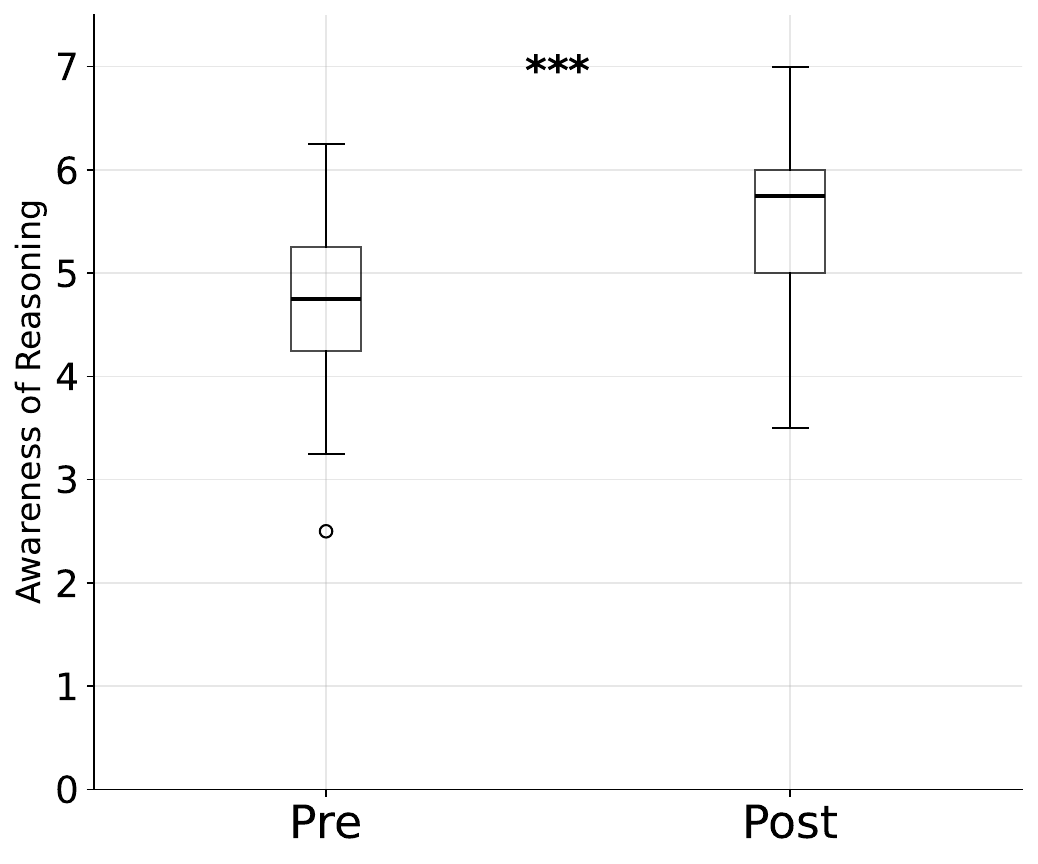}
        \Description{Box plot of ethical consequence reasoning scores on the y axis and time points (pre/post) on the x axis}
        \caption{Ethical consequence reasoning}
    \end{subfigure}
    \caption{Changes in ethical sensitivity in technology design scale scores. Statistical significance between pairs is represented via asterisks (*** $p < .001$).}
    \label{fig:RQ1_Result}
    \Description{Two side-by-side box plots comparing pre vs. post scores on a 0-7 scale. Panel (a), Ethical issue awareness, shifts upward from around the mid-5 range at pre to around the low-6 range at post. Panel (b), Ethical consequence reasoning, shifts upward from around the high-4/low-5 range at pre to around the mid-5 range at post. Asterisks indicate a significant pre-to-post increase (p < .001).}
\end{figure}

The two sub-constructs---ethical issue awareness and ethical consequence reasoning---demonstrated acceptable internal consistency in the pre-survey, as indicated by Cronbach’s $\alpha$ values of .709 and .812, respectively, a reliability coefficient commonly used to assess the internal consistency of scale items, with values above .70 generally considered acceptable~\cite{tavakol2011making}.
Results from the pre-post paired $t$-tests demonstrated statistically significant gains across both sub-constructs of ethical sensitivity ($p$ < .001), with large effect sizes (Cohen’s $d$ = 1.43 for awareness, Cohen’s $d$ = 1.28 for reasoning), as demonstrated in Fig. \ref{fig:RQ1_Result}. As a robustness check, we additionally conducted Wilcoxon signed-rank tests, which reaffirmed the paired $t$-test results (ethical issue awareness: $p<.001$, matched-pairs rank-biserial $r=.99$; ethical consequence reasoning: $p<.001$, $r=.98$).  Participants appreciated the opportunity to engage in ethical discussions and learn diverse perspectives. For instance, G4P1 noted, \emph{``Opportunities like this are rare. Just having the chance to discuss ethics makes me more aware of ethical issues. (...) Since the discussion is well-structured, it also feels easier to learn different considerations.''}

Between two sub-constructs, the greater improvement was observed in the awareness, suggesting that participants became substantially more attuned to potential ethical issues than to reasoning about the consequences of a design decision. Specifically, participants credited the \textit{Identify Stakeholders} and \textit{Write Reviews} stages with encouraging them to consider ethical issues from the perspectives of multiple people and situations they had never been aware of. They noted that they not only received the most challenges to their usual way of thinking but also gained the most in terms of ethical awareness during these stages. For example, G13P3 stated, \emph{``As an engineering student, I usually tend to go straight into the problem and solution, so going through the Identify Stakeholders stage at the beginning felt new. I was able to broaden my perspective.''}

These findings suggest that participating in the group ethical discussion activity enhances ethical sensitivity by supporting a multidimensional awareness of ethical issues, expanding the range of ethical dimensions participants considered when evaluating design decisions.

\subsection{RQ2. Facilitating Structured Ethical Discussions}
Participants expressed that the discussion structure (i.e., divergence, groan, and convergence), the LLM facilitator, and the LLM participants made the stages easy to understand and follow, reducing the need for additional instructions or an external intervention.

\subsubsection{Familiar, Structured Flow for AI-Supported Discussion}
\label{lab:discussionstructure} %
The discussion structure was regarded as appropriate for covering and understanding a variety of ethical perspectives within a limited time.
As we structured the discussion following Kaner et al.~\cite{kaner2014facilitator}, participants reported that the phase composition, comprising the divergence, groan, and convergence phases, was sufficiently simple and resembled a typical human discussion sequence of opinion-question. As G9P2 noted, \emph{``Isn’t this procedure quite similar to that of a typical discussion? I don’t think discussing with AI makes it particularly more difficult.''}

They also found the discussion structure suitable for brainstorming, inviting free expression without pressure in the \textit{Idea-gathering} phase and leading to the expansion of the discussion in the subsequent phases. G3P2 stated, \emph{``Rather than being a system that goes all the way end-to-end, I think its real value lies in the brainstorming stage. In brainstorming, it is important for everyone to be able to share their ideas comfortably, and in that respect, I felt that having an idea-gathering phase was particularly meaningful.''}
In addition, participants described the stacking-based format as enabling parallel discussion of diverse ideas. However, some participants also noted that the format limited opportunities for in-depth discussion on a single idea.
G8P2 noted, %
    ``\emph{The discussion proceeded in a way where each person spoke in turn, and other participants couldn't intervene. While it was suitable to talk about a variety of ideas, it was difficult to explore them in depth.}''

\subsubsection{LLM Facilitator Maintaining Pace and Coherence} %
Overall, participants reported that by maintaining pace and coherence, the LLM facilitator enabled them to complete the discussion smoothly within the allotted time. Participants positively described the facilitator’s role in the discussion, such as speaking turn control, time management, and content summarization. %
They reported that, by managing speaking turns and time spent in each phase, the LLM facilitator provided a sense of reassurance that the discussion was progressing as intended. By keeping phases on schedule, the facilitator gave participants confidence that the discussion was on track, which made the session feel more deliberate and task-focused and helped them sustain attention.
G1P1 stated, \emph{``Unlike ChatGPT, which can often feel disorganized or make it hard to see the progress of ideas, a feeling that the facilitator was coordinating discussion flow gave me more of a sense of engaging in an actual discussion, which made it easier to immerse myself.''}

However, some participants expressed discomfort with the facilitator’s restrictions on speaking turns. While they agreed that a certain level of control is inevitable given the nature of discussions involving multiple participants, they suggested more flexibility, such as ``\emph{allowing partial interruptions during turn}'' (G10P1).%

Participants also reported that the facilitator's summarization reduced their cognitive burden. They mentioned that keeping up with the conversation flow was sometimes challenging due to factors such as the AI’s quick pace, the return of past discussion points, or the need to organize their own thoughts. In these cases, being able to catch up by referring to the summaries was reported to be helpful. G13P2 stated, \emph{``Since the AI spoke so quickly, things passed by before I could fully process them, but the summaries helped me keep up.''} Notably, participants emphasized the advantage of the facilitator being AI-based. G13P1 noted, ``\emph{If it were a human, providing summarization might take a long time, and the flow would have dragged. In that case, it probably wouldn’t have been as helpful.}''

\subsubsection{LLM Participants Guiding Users to Stay on Topic}
While the LLM facilitator helped participants maintain and follow the discussion structure, participants reported that the LLM participants were helpful in clarifying the purpose of each stage and maintaining the focus of the discussion. They said LLM agents' opinions were well aligned with the discussion goals and made it easier for participants to understand what was expected of them. G2P3 noted, \emph{``In the stakeholder stage, the AI gave me ideas that helped me concretely imagine stakeholders. At first, when asked to discuss stakeholders, I thought ‘everyone indirectly affected is a stakeholder’ and wrote that down. But then the AI suggested something specific like ‘a pedestrian walking on the street,’ which helped me get a clearer sense of what was meant.''}

The LLM participants also played a role in keeping discussions on track when other participants went off-topic. Since the LLM agents were prompted to produce only topic-related opinions and questions, participants were able to quickly return to the main subject even when their contributions diverged or their thoughts wandered.
For example, when it was easy to prematurely discuss problems during the stakeholder stage or propose solutions during the problem stage, the LLM agents remained on-topic, helping participants stay focused on the specific goals of each stage. G13P3 noted that, unlike in human-only discussions, it was beneficial not having to make the effort to steer the conversation back, \emph{``Since it was a 3:3 environment, even if the other two participants strayed, I didn’t have to handle it. (...) Because AI agents focused on the topic regardless of what they said, others also steered the conversation back on track.''}

In addition, participants reported that the agents’ questions prompted deeper ethical reflection, encouraging them not only to clarify their perspectives but also to reflect more critically on their own views. As G5P1 explained, the LLM agents often asked \emph{``the kinds of questions that people might want to clarify but would feel uncomfortable asking themselves because the questions seem too obvious or general.''} For instance, when G5P2 mentioned \emph{``the Memora data manager''} as a stakeholder, G5P1 assumed this was straightforward from a data protection perspective and therefore refrained from probing further. However, when the AI asked a clarifying question, \emph{``Could you elaborate on what you mean by the data manager?''}, G5P2 clarified that \emph{``a data manager might attack user privacy by misusing information through unauthorized access,''} thereby enabling G5P1 to better understand G5P2’s intention. In a later interview, G5P2 also revealed that he had mentioned the data manager without much thought at the time, yet the AI’s intervention prompted them to reflect and articulate a more specific concern about potential misuse of user information. %

\subsubsection{LLM Participants Facilitating Diverse Perspective Taking}
Participants reported that, although their groups consisted only of STEM students, the LLM agents with distinct ethical viewpoints enabled them to encounter a wide range of ethical perspectives beyond what STEM only groups can offer. For example, G4P1 noted, \emph{``Since the AI agents were assigned diverse fields of expertise, the discussion could be approached from multiple perspectives--such as social, technological, governmental, and security aspects. If the discussion had been only among STEM students, such ideas would not have emerged.''}

Moreover, participants highlighted that LLM agents consistently represented minority or less dominant viewpoints without the social pressures that human participants might experience. As G13P2 explained, \emph{``If this had been an actual classroom, I think the welfare-major students might have felt discouraged, since the engineers [constituting the majority] wouldn't really listen to them. But with AI, that kind of issue didn’t arise, so it was helpful.''} Also, participants described the consistent presence of these perspectives as a reminder that shaped their own reasoning. For example, G7P2 noted, \emph{``Since the care ethics agent consistently brought up issues related to marginalized people, I think I unconsciously searched for solutions while wondering, ‘Would [the care ethics agent] Yeon-Su Song be satisfied with this answer?''}

However, participants also noted that while this diversity supported broad exploration, it sometimes came at the expense of depth, highlighting a trade-off between breadth and depth within structured ethical discussions. G15P1 reflected, \emph{``Because the AI-generated ideas for each specific domain, the discussion tended to develop in a divergent way.''}

\subsection{RQ3. Participants' Assessment of LLM Agents' Contributions}

\begin{figure}[t]
    \centering
    \begin{subfigure}[t]{0.32\textwidth}
        \centering
        \includegraphics[width=\linewidth]{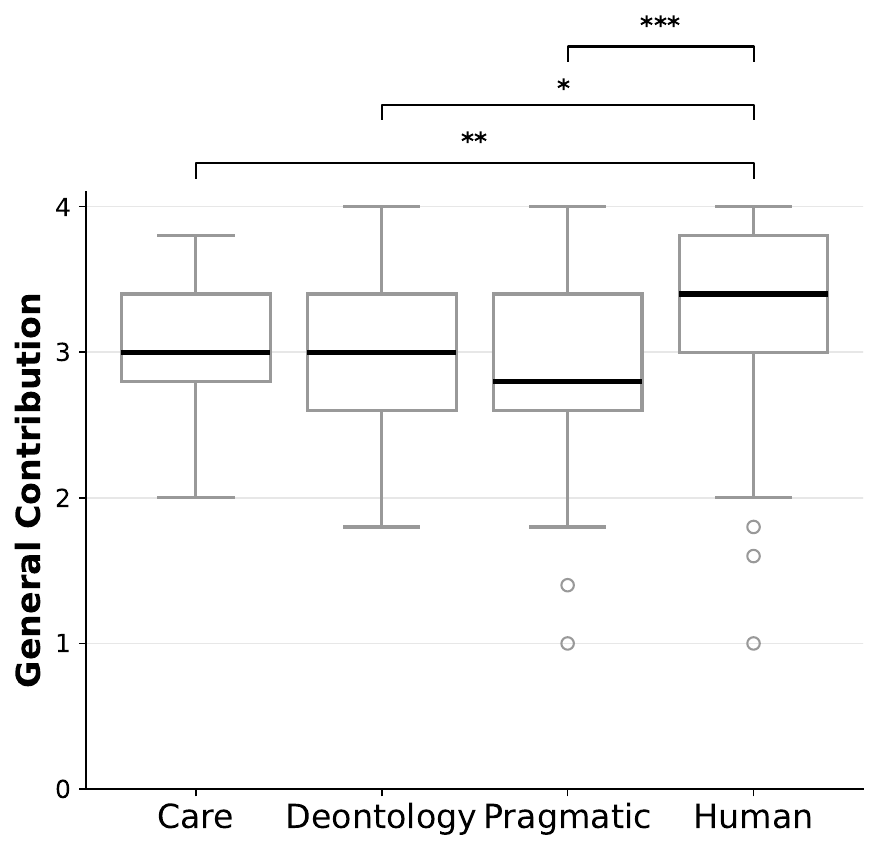}
        \Description{Box plot of awareness of Stakeholder perspective scores on the y axis and pre-/post-comparison on the x axis}
        \caption{General Contribution}
    \end{subfigure}
    \begin{subfigure}[t]{0.32\textwidth}
        \centering
        \includegraphics[width=\linewidth]{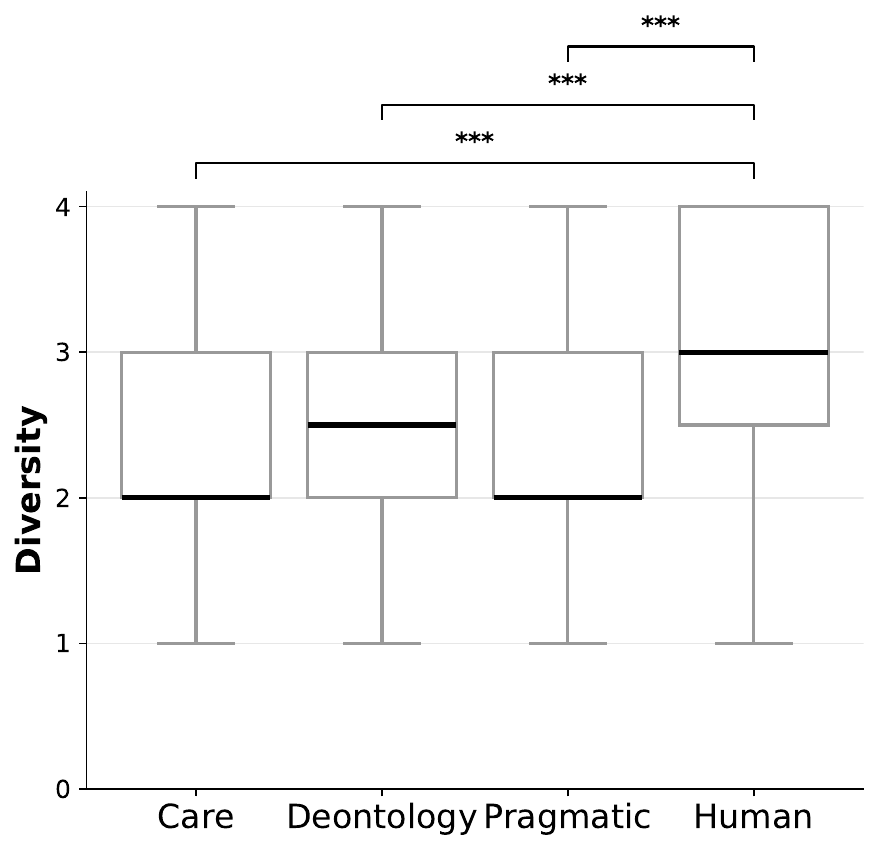}
        \Description{Box plot of awareness of bias scores on the y axis and pre-/post-comparison on the x axis}
        \caption{Diversitiy}
    \end{subfigure}
    \begin{subfigure}[t]{0.32\textwidth}
        \centering
        \includegraphics[width=\linewidth]{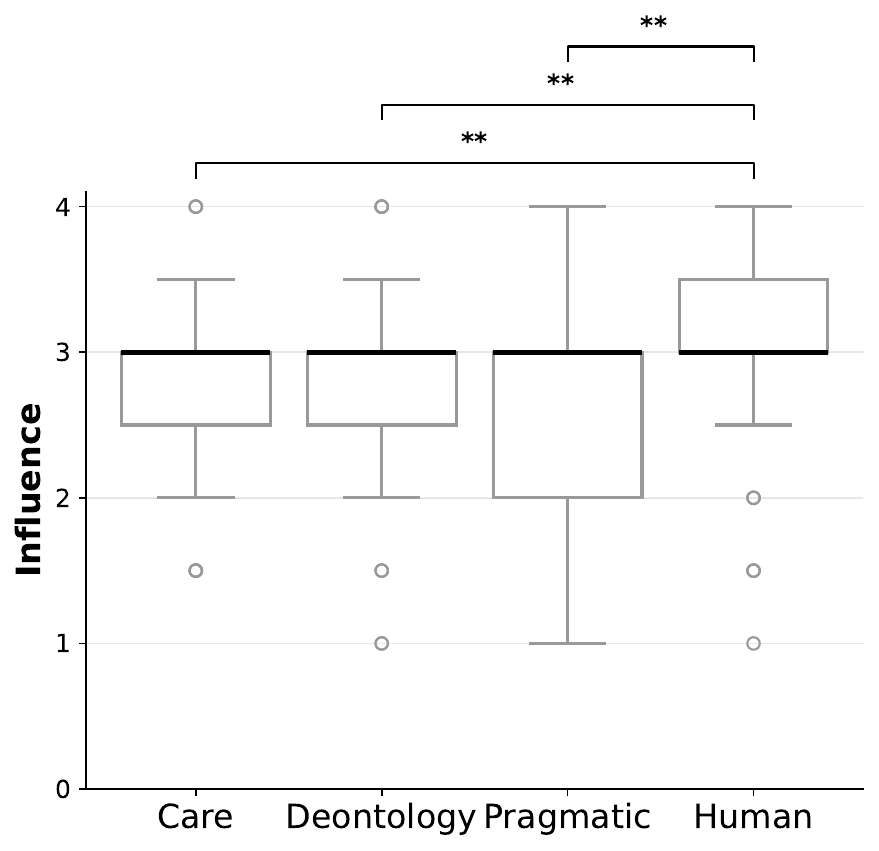}
        \Description{Box plot of awareness of bias scores on the y axis and pre-/post-comparison on the x axis}
        \caption{Influence}
    \end{subfigure}
    \caption{Peer evaluation scores across the four participant groups (Care, Deontology, Pragmatic agents, and Human peer) in three dimensions: (a) General Contribution, (b) Diversity, and (c) Influence. Statistical significance between groups is represented via asterisk (* $p < .05$, ** $p < .01$, *** $p < .001$).}
    \Description{Figure 4 presents three boxplots showing peer evaluation scores for four participant groups—Care agent, Deontology agent, Pragmatic agent, and Human peer—across the dimensions of general contribution, diversity, and influence. The vertical scale ranges from 0 to 4. Overall, the AI agent groups cluster between 2 and 3 points, while the Human peer group consistently falls between 3 and 4 points. In all three dimensions, Human peers show higher median scores than agents, and these differences are statistically significant, indicating more positive evaluations of Human peers compared to AI agents.}
    \label{fig:RQ2_Result}
\end{figure}

As shown in Figure \ref{fig:RQ2_Result}, we conducted   Kruskal--Wallis tests to examine group differences among participants across three peer evaluation dimensions: general contribution, diversity, and influence. Across all three dimensions, human peers were rated significantly more positively than  LLM agent peers($p<.05$ for all pairwise comparisons against each LLM persona). In contrast, no significant differences were observed among the different LLM personas (care, deontology, and pragmatic personas). Detailed statistics for each dimension are reported in the Appendix (Table~\ref{tab:peer_rating_analysis}). Overall, participants’ experiences revealed a gap between their expectations of LLM agents as reasoned discussants and credible experts and the agents’ actual behavior during discussion.

\subsubsection{Overly Accepting Responses Limiting Debate Depth}
\label{sec:overlyaccepting}
Participants reported that the LLM agents frequently exhibited sycophantic behaviors, which led them to rate the agents’ general contribution lower. Participants expected that LLM agents would refute and oppose ideas to better engage with ethical discussion.
However, unless faced with a question that was entirely wrong or one that outright denied certain values (e.g., \emph{``I think engineers do not have to care about ethics,''} and \emph{``Fairness is not what we have to concern,''}) the LLM agents tended not to refute but rather to accept the point. Because of this, users reported that the discussions felt less competitive compared to human-to-human debates. %
While less competitive discussions made participants engage in an emotionally comfortable environment, participants reported that, because the LLM agents appeared to agree with and show interest in every question and opinion raised by human participants, regardless of their assigned personas, they received fewer benefits from the personas of the agents in the later stages of the discussion.
Participants highlighted that one advantage of LLM agents was that receiving questions or counterarguments from LLM agents caused less emotional impact than receiving them from human peers. However, they said that this advantage was not fully utilized in the discussion. They reported that the lack of rebuttals and rigorous argumentation limited the extent to which deeper ethical thinking could be fostered, which ultimately made participants rate LLM agents' general contributions lower.%

\subsubsection{Desire to Understand Agents’ Ways of Thinking}
Participants reported that the agents’ lack of reasoning undermined the perceived diversity of the agents, leading to a lower diversity rating than that of human peers. Participants expected more than just consistency and repetitive, patterned answers from LLM agents; instead, they sought detailed expert reasoning combined with distinct persona identities, so they could better contrast perspectives and advance their ethical reasoning.
While participants acknowledged that they encountered ideas they had not previously considered, they did not necessarily perceive them as unique or substantially different from their usual ways of thinking. In other words, even if the agents presented ideas that were somewhat difficult to think of, if they were easy to accept, participants tended to dismiss them without much attention. G14P3 highlighted, \emph{``The AIs did present ideas that reflected their traits and majors, but it felt like those ideas didn’t really connect to deeper ethical thinking. It was more like, ‘oh, okay,’ and then just moving on.''}

Participants expressed that rather than simply receiving the agents’ ideas or keywords, they wanted to understand how the agents’ ways of thinking differed from their own and to use that contrast to advance their ethical reasoning. They emphasized that they were more interested in understanding why the AI generated certain ideas and why it considered them important, rather than the ideas themselves. G10P2 noted, \emph{``It feels like the conversation is staying in the same place. We only talk about the opinions, but not the reasoning behind them, so it feels like we’re just looking at the trees and not the forest.''}
Participants expressed that the development of ethical reasoning requires a holistic understanding of how an agent thinks as an entity. They said receiving diverse ideas alone was not sufficient to foster diverse perspectives. They wanted to draw clear images of how agents thought. However, since the agents did not demonstrate such depth, participants said that the AI played only a lubricating or supporting role in the discussion, which led them to rate the LLM agents' diversity lower.

In line with this, participants wanted to learn the agents’ ways of thinking from a broader perspective. They suggested features to better distinguish the distinct 'voices' of individual agents, enabling them to recognize who said what, track each agent’s contributions, grasp their identity and reasoning process, and review their reasoning processes collectively. For instance, G6P3 suggested, \emph{``A UI where I can see each agent's dialogue at a glance is needed. That way, I can easily review the AI’s opinions in one place and understand the overall trends.''}

\subsubsection{Ambivalent Expectations for LLM agents as Both Generic LLM and Persona Expertise}
Participants reported that    either minor errors or generic ideas from LLM participants quickly led to disappointment and made them less receptive to the agents. They wanted LLM agents to perform both flawless, knowledge-rich roles expected just as a general LLM, as well as the experience-grounded roles associated with their persona's domain expertise.

Participants wanted the AI to adopt methods similar to those used by LLM-based chatbots (e.g., ChatGPT, Gemini) to ensure reliability. They expected the agent to present its ideas with references to sources such as online articles or academic papers. When asked why they expected this from the agent but not from human participants, one participant explained that, based on prior experiences with chatbots providing information along with its sources, they viewed human discussants as collaborators who could learn together when uncertain, whereas they expected AI to deliver ideas accurately and with proper references. Therefore, participants reported that when agents showed flaws or offered claims without references, their perceived reliability dropped sharply, much more than it did for human discussants. As G3P2 noted, \emph{``ChatGPT often backs up its answers with references or URLs, and I expected the agents to do that too. (...) Without references, I cannot trust it.''}

Also, participants tended to be stricter when evaluating the credibility of LLM participants, viewing implausible statements from the agents as reflecting AI's limitations rather than the characteristics of their personas as discussants.
For example, when the care ethics agents with a social welfare background, without an engineering background, made a technically impossible claim, participants perceived it as an AI flaw and reported a decrease in trust, rather than viewing it as an acceptable knowledge gap for that persona. Specifically, when this agent suggested inserting an LED into a contact lens, G3P3 thought, \emph{``Inserting an LED into a lens doesn’t make sense,''} which led to reduced trust. Interestingly, in the follow-up interview, G3P3 remarked that if a human social welfare major had proposed the same idea, they would not have seen it as a credibility issue, but rather thought, \emph{``Since they’re not an engineer, that’s understandable.''} and would likely have explained why the idea was technically infeasible.

However, participants were also disappointed when the agents failed to show the unique insights expected from their personas. While participants wanted agents to perform the benefits of AI, they did not presume that the agents’ lack of personal experiences, insights, or reasoning styles---traits expected from their persona descriptions---was simply because AI cannot have personal experiences or insights. Instead, they expected the agents to demonstrate both the vast knowledge of AI, including engineering expertise, and the personal experiences and insights associated with their assigned personas. In other words, participants did not consistently perceive the agents either as an LLM-based chatbot like ChatGPT or as entities embodying specific personas; rather, depending on the point at which their expectations were unmet, they appeared to alternate between an “LLM heuristic” and a “domain-expertise heuristic” in interpreting the agents’ behavior. As G10P3 stated, \emph{``Since it’s AI, I assumed it would have a lot of knowledge. So I wanted it to quickly provide sources as well. (…) I also expected it to demonstrate expertise just as I would expect from a philosophy or social welfare student.''}

\subsection{RQ4 Social Dynamics in Multi-Human, Multi-Agent Ethical Discussion}
This section examines how multi-human, multi-agent ethical discussions shaped social dynamics, including participation pressure, interaction patterns, and participants’ attitudes toward LLM agents.

\subsubsection{Reduced Social Pressure, but Limited Human--Human Interaction} %

Participants expressed experiencing \emph{lower social pressure} in the mixed human-AI environment. They reported being able to speak more directly to LLM agents because they perceived LLM participants as lacking emotions; this allowed them to focus more on logical structure and precise wording rather than worrying about hurting the agents. Participants emphasized that it was important to express their genuine thoughts freely, and the multi-agent structure helped them to do so.
Even though participants knew that other human participants were also observing their questions and answers, they noted that the sense of “debating with LLM agents one-on-one" helped them express their opinions more honestly, making them feel less concerned that their conversations were visible to others.%

Furthermore, participants did not feel pressured to contribute at every moment and could instead focus on organizing and refining their own ideas.
This was because the discussion could progress even without constant input from human participants.
They reported not experiencing the typical pressure to fill silences that often arises in ordinary discussions, and this allowed them to concentrate on improving the quality of their contributions.
Nevertheless, participants reported that seeing others actively participate created a sense of obligation to contribute, which sustained healthy pressure to share new ideas when they felt they had something meaningful to add.

On the other hand, the human-AI mixed ethical discussion environment \emph{reduced human-to-human exchanges}. During the discussion, participants frequently directed clarifications or critiques to the LLM participants rather than to the human participants, decreasing human--human communication. This pattern is notable given our findings from RQ3. While participants rated AI contributions lower and acknowledged human responses as more valuable and useful in interviews, participants still felt it was easier to interact with the AI. Participants mentioned that asking questions to the AI involved less emotional burden, whereas they were more cautious when asking follow-up questions to human participants because it could affect relationships in the group, worried that their questions might come across as confrontational, and they avoided these relational concerns by addressing the AI instead. %
As G10P3 noted, \emph{“When the AI expressed an opinion similar to a person’s, and I had a question about that opinion, I found it easier to ask the AI rather than the person. But in doing so, I only received about 80\% of the answer I wanted, not the full 100\%.”}
The distribution of questions in the logs points the same way. Participants addressed 79.7\% of their questions to LLM agents (59 of 74), well above the 60\% their availability predicts (Wilcoxon signed-rank against .60, M = .81, SD = .29, p = .017, r = .72).

\subsubsection{Respect for LLM agents in Mixed Human-AI Engagement} %
Participants reported that the mixed environment of multi-agent and multi-human fostered a more respectful attitude toward LLM agents.
They noted that if they had discussed only with LLM agents, without other human participants, they would likely have dominated the interaction, try to control agents as they want.
Their anticipated behaviors without the presence of other human participants included treating LLM agents as subordinates or mere tools to command, dismissing ideas that did not align with their own views or interests, or directing the agents to focus only on the ideas they personally cared about---thus making it difficult to move beyond the participant's own ethical perspective. As G6P1 reflected, \emph{``When only AI is present, I feel like I can freely set the direction. I think I would be tempted to keep giving commands like, `I'm not interested in it. Try discussing from a completely new perspective.'''}
In contrast, within the mixed environment of multi-agent and multi-human discussion, the presence of other human participants led them to treat the AI’s contributions with greater respect.
Participants explained that they read the AI’s opinions more carefully, engaged in refuting or questioning them, and in doing so, the discussion progressed in a way that was less biased and not overly tailored to a single participant or perspective. %

\section{Discussion}

\subsection{Beyond Ideas Toward Ways of Thinking: Learning Ethical Reasoning Through Multi-Agent AI Interactions}
Our findings suggest that participants wanted to learn ethical reasoning which agents try to represent, rather than simply collecting new ideas. Beyond simply gaining ideas they had not considered, participants wanted to treat the agents as ethical mentors as well as peers and learn how agents conduct ethical reasoning. This aligns with prior work in two ways. Tran et al.~\cite{Tran2024Student} found that STEM students desire ethical consultants in ethics education who can help them develop ethical awareness, while Chan and Hu~\cite{chan2023students} reported that students expect AI to provide constructive feedback beyond mere checking or brainstorming, enabling them to improve their skills through AI. In the current system, personas were defined mainly by demographic characteristics and a primary ethical perspective, with prompts oriented toward producing stage-specific ideas. While effective for generating diverse ideas, this design was not sufficient for helping participants  understand agents' ethical reasoning frameworks.

Along these lines, participants emphasized that they wanted to  track each agent's discussion activities as a way to better grasp the agents’ output pattern. They reported that, in order to understand an  agent's ethical perspective, they wanted to see all of that agent's turns together rather than interleaved with other participants' contributions. Aligned with prior research \cite{ha2024clochat}, these findings suggest that the discussion structure should be designed to help participants easily gather and recall each agent’s viewpoints by providing distinctive traits, so that they can integrate these perspectives into a coherent ethical framework and actively learn from the agents’ reasoning processes.

Building on this perspective, our findings highlight the need to shift agent outputs from being conclusion-oriented to more process-oriented. By leveraging the persona’s values and synthetic experiences, agents could surface the intermediate steps of deliberation, enabling participants to learn not just what ethical judgments are made, but also how they are formed.

However, this recommendation should be applied cautiously. Prior work warns that anthropomorphizing an AI as an authoritative expert fosters over-reliance and erodes learner agency and questioning~\cite{qadir2026psychology}. Exposing agents' reasoning could amplify this authority, since fluent reasoning may read as expertise. To mitigate this risk, we designed the agents with undergraduate student personas rather than expert ones and included a dedicated Questioning phase where people can challenge the AI. Process-oriented agent design should therefore be paired with such safeguards that frame agents as fallible peers whose reasoning invites critique.

\subsection{Managing Expectation and Identity Disclosure to Mitigate AI Agent Dismissal}
Our results reveal that a critical challenge in role playing based ethics education using AI agents: when participants encountered errors  in agents, they tended not to interpret errors as natural boundaries of a persona’s domain expertise, but rather as defects of generative AI itself.
This dismissal tendency is concerning because prior research has shown that the value of role playing lies not only in exposing learners to different ethical considerations but also in helping them practice communication strategies with diverse stakeholders, including those who may have incomplete reasoning or biased viewpoints~\cite{May2014Influence, Castro2023Piloting}. Also, research on human discussion dynamics further suggests that ignoring participants can have detrimental effects~\cite{zhang2013structural}, limiting both the diversity and depth of deliberation. In contrast, participants in our study expected AI agents to function as flawless entities, and when flaws became visible, they quickly discredited the agent’s persona. This did not appear to reflect a misunderstanding of the agents’ roles, which were clearly structured by discussion stage and explicitly presented as distinct facilitator and persona-based participant roles. Rather, participants evaluated these role-based agents against expectations formed through prior interactions with general-purpose AI tools~\cite{lenskjold2023should, cai2025understanding}. Such expectation carryover may also arise more broadly in multi-agent systems.

Prior research has shown that disclosing the functional limitations of AI can enhance user satisfaction in high-contact services and reduce negative reactions when errors arise~\cite{Mozafari2020Chatbot, Mozafari2021Resolving,Yu2024Who}. Building on these works, expectation setting can be beneficial in the context of role playing based ethics education. Onboarding procedures can transparently communicate the limitations of agents' personas to participants, framing these constraints as part of their persona and learning experience rather than as unexpected failures. For example, an AI role playing a welfare specialist could state at the outset: \emph{“I may overlook technical details, but I can contribute insights into social and institutional impacts.”} Such disclosure can not only calibrate user expectations but also reposition engagement with agent shortcomings as part of the learning process.

Another approach is to experiment with the timing of disclosure. While pre-session disclosure can build transparency, post-session disclosure, where participants do not know which discussants are AI, may reduce bias and foster more authentic interaction~\cite{mozafari2022trust}. However, this carries the risk of participants inferring agent identities from errors, which may lead to feelings of deception. In multi-human, multi-agent environments such as ours, a hybrid strategy may be most effective: first, disclosing the presence and general constraints of AI agents pre-session, and then revealing which participants were AI agents at post-session. Future work should investigate how such strategies influence learners’ trust, respect, and willingness to engage with AI agents as learning partners.

\subsection{Designing for Balanced Social Dynamics in Mixed Human--AI Ethical Discussions}
Our findings highlight a trade-off in introducing AI agents. While they lower social barriers to participation, they may do so at the cost of human-human interaction. Agents' overly agreeable responses and their lack of lived experience to support their ideas rendered discussions shallower and less critically challenging.

We argue that one way to deal with this issue is to improve the response strategy of AI agents. Rather than offering universally supportive replies, AI agents could be designed to raise counterarguments, challenge assumptions, or highlight trade-offs, thereby encouraging deeper ethical reasoning~\cite{Lee2025Conversational}. Additionally, agent personas can be enriched with more vivid narratives, motivations, and even simulated experiences~\cite{park2023generative} to compensate for the lack of personal experiences. However, strengthening personas and giving AI greater autonomy also introduces challenges. Prior research shows that when AI appears excessively autonomous and simulates experience, users may perceive it as less trustworthy~\cite{Wen2025Trust}. To address this risk, transparency becomes essential: systems should explicitly communicate the AI’s role and limitations to maintain credibility~\cite{mozafari2022trust}. In addition, as Wen et al.~\cite{Wen2025Trust} suggest, giving users direct control over the factuality or assertiveness of AI responses may help strike a balance between richer persona design and sustained trust.

We also suggest that encouraging deeper human--human interaction could be part of the solution.
Structural mechanisms can be introduced to prioritize human engagement. For example, we could request participants to respond to at least one human peer before turning to the AI, or provide feedback and relational indicators that highlight the value of peer contributions. Such features would remind students of the unique benefits of human perspectives, particularly the sharing of lived experiences and subjective insights that AI cannot replicate. Also, we could explicitly disclaim that LLM agents lack personal experiences and thus encourage students to turn to human peers when human--human interaction is limited~\cite{bo2024disclosures}. By embedding these mechanisms into the discussion design, systems can avoid overreliance on AI and sustain the collaborative and dialogical nature of ethical learning.

\subsection{Limitations and Future Work}

Using LLM agents in ethics education requires careful consideration of how ethical perspectives are represented. We explicitly specified each persona’s roles, values, and opinions and validated their consistency through simulated discussions, thereby constraining the range of possible responses and reducing unintended influence from the underlying model. However, these perspectives necessarily reflected our design choices and could not represent all relevant ethical positions. We therefore disclosed the agents’ identities, positioned them as peers, allowed students to challenge their contributions, and maintained human oversight. Nevertheless, fluent but unsupported responses may still acquire undue authority, and these safeguards cannot fully eliminate model bias or overreliance. LLM agents should therefore be treated as contestable prompts for deliberation rather than sources of a ``correct'' ethical answer.

Beyond these ethical concerns, our study has several methodological and empirical limitations. Our evaluation relied solely on pre--post comparisons without a control condition, which limited causal interpretation of the observed effects. Without a control or comparison group, it is difficult to disentangle the impact of the system from other factors such as novelty effects or general learning effects. However, constructing a control condition in this context was not straightforward. Comparing our discussion system with previous lecture based education cannot disentangle our results from a novelty effect. Comparing our 3:3 human--AI setting against a 3:0 human-only setting would confound the effect of AI presence with the effect of simply changing the number of participants, while comparing 3:3 against a 6:0 all-human setting would additionally confound configuration of each AI.
Without established guidelines for configuring multi-agent, multi-human discussions, we could not ensure a fair comparison. For example, agents designed to speak too long, too fast, and too often could dominate discussions and reduce human critical thinking~\cite{Song2024Multi}. On the other hand, agents designed to speak less or more slowly may deflate learning gains that AI agents can introduce. Future work should identify a ``fair'' agent configuration and pursue controlled comparison studies that can better isolate the effect of different dimensions of AI participation from these confounds.

Our results may also depend on how the agents are configured and implemented. For example, equipping agents with retrieval-augmented generation (RAG) to ground their claims in citable sources could increase participants' trust in the agents and lead to more positive evaluations. Similarly, we designed our system around a frontier model (GPT-4o) at the time of this study; because we specified prompting for each agent's behavior including utterance length, speaking style, and tone, we expect that other frontier models would produce similar interaction patterns and results. However, as frontier models continue to evolve rapidly and a growing variety of LLMs become available, future work should systematically verify whether our findings hold across different models. In sum, our study demonstrates that multi-human, multi-agent discussion can be beneficial for ethics education. However, different model choice, prompting strategy, and memory or retrieval architecture may shape depth and quality of agent responses %
 and change participants' perceptions and learning outcomes, which should be studied systematically in future work.

In addition, our study focused on short-term changes measured immediately after the intervention, limiting us unable to observe sustained improvements. Prior research suggests that short-term gains in ethics training do not necessarily translate into long-term effectiveness~\cite{kreismann2021business}. Future studies should include longitudinal designs to assess whether gains in ethical reasoning and sensitivity persist over time.

Finally, participants within groups were strangers to one another and entered the discussion without prior knowledge. In real-world technology design projects in university classes, however, participants may bring pre-existing knowledge or established relationships with peers or mentors, which could alter group dynamics~\cite{senior2014learning}. Recent work has also suggested that the use of generative AI may erode social learning~\cite{hou2025all}, including possible negative effects on mentorship, peer learning, and overall motivation. While our focus in this system was to leverage LLM agents to scaffold novice designers' initial engagement in group discussions with other students on ethical issues, we caution against relying solely on LLM agents as a sustained substitute for peer discussion. Future research should test  systems like this one in authentic educational contexts to better understand and mitigate these potential risks.
\section{Conclusion}
This paper presented Ethics Training Agents, the first multi-agent group discussion system designed to facilitate collaborative ethics training for STEM students. Our study with 45 participants showed that while the system enhanced ethical sensitivity, sustained engagement, and supported perspective-taking in group discussions, it also revealed social dynamics in the mixed human--AI environment. The findings further suggested design implications for future multi-agent discussion systems: consistent facilitation and topic alignment can effectively promote coordination and reflection, yet sustaining richer debate dynamics may require agents that adopt less agreeable stances and provide more diverse argumentative strategies. This study also identified participants' ambivalent expectations toward agents that combine vast knowledge and resources with distinct persona identities to foster deeper ethical reasoning. Overall, this work demonstrated the promise of mixed multi-agent and multi-human discussions for scalable ethics training and offered practical insights for designing future education systems.

\bibliographystyle{ACM-Reference-Format}
\bibliography{bib.bib}

@inproceedings{hou2025all,
  title={'All Roads Lead to ChatGPT': How Generative AI is Eroding Social Interactions and Student Learning Communities},
  author={Hou, Irene and Man, Owen and Hamilton, Kate and Muthusekaran, Srishty and Johnykutty, Jeffin and Zadeh, Leili and MacNeil, Stephen},
  booktitle={Proceedings of the 30th ACM Conference on Innovation and Technology in Computer Science Education V. 1},
  pages={79--85},
  year={2025}
}

@article{harpe2015analyze,
  title={How to analyze Likert and other rating scale data},
  author={Harpe, Spencer E},
  journal={Currents in pharmacy teaching and learning},
  volume={7},
  number={6},
  pages={836--850},
  year={2015},
  publisher={Elsevier}
}

@inproceedings{qadir2026psychology,
  title={The psychology of learning from machines: anthropomorphic AI and the paradox of automation in education},
  author={Qadir, Junaid and Mumtaz, Muhammad},
  booktitle={2026 IEEE Global Engineering Education Conference (EDUCON)},
  pages={1--10},
  year={2026},
  organization={IEEE}
}

@article{tavakol2011making,
  title={Making sense of Cronbach's alpha},
  author={Tavakol, Mohsen and Dennick, Reg},
  journal={International journal of medical education},
  volume={2},
  pages={53},
  year={2011}
}

@inproceedings{hwang2023aligning,
  title={Aligning language models to user opinions},
  author={Hwang, EunJeong and Majumder, Bodhisattwa and Tandon, Niket},
  booktitle={Findings of the Association for Computational Linguistics: EMNLP 2023},
  pages={5906--5919},
  year={2023}
}

@article{jagger2011ethical,
  title={Ethical sensitivity: A foundation for moral judgment},
  author={Jagger, Suzy},
  journal={Journal of Business Ethics Education},
  volume={8},
  number={1},
  pages={13--30},
  year={2011},
  publisher={Philosophy Documentation Center}
}

@inproceedings{pierrakos2019reimagining,
  title={Reimagining engineering ethics: From ethics education to character education},
  author={Pierrakos, Olga and Prentice, Mike and Silverglate, Cameron and Lamb, Michael and Demaske, Alana and Smout, Ryan},
  booktitle={2019 IEEE Frontiers in Education Conference (FIE)},
  pages={1--9},
  year={2019},
  organization={IEEE}
}

@article{braun2006using,
  title={Using thematic analysis in psychology},
  author={Braun, Virginia and Clarke, Victoria},
  journal={Qualitative research in psychology},
  volume={3},
  number={2},
  pages={77--101},
  year={2006},
  publisher={Taylor \& Francis}
}

@misc{cmu_groupwork,
  author    = {Eberly Center},
  title        = {Peer Evaluation Form for Group Work},
  howpublished = {Carnegie Mellon University},
  year         = {n.d.},
  url          = {https://www.cmu.edu/teaching/designteach/design/instructionalstrategies/groupprojects/tools/index.html},
  note         = {Accessed: 2025-06-16}
}

@article{Hess2018Systematic,
    author    = {Justin L. Hess and Grant Fore},
    title     = {A Systematic Literature Review of US Engineering Ethics Interventions},
    journal   = {Science and Engineering Ethics},
    volume    = {24},
    number    = {2},
    pages     = {551--583},
    year      = {2018},
    month     = apr,
    doi       = {10.1007/s11948-017-9910-6}
}

@inproceedings{Hanschke2024Data,
    author    = {Victoria A. Hanschke and David Rees and Merve Alanyali and others},
    title     = {Data Ethics Emergency Drill: A Toolbox for Discussing Responsible AI for Industry Teams},
    booktitle = {Proceedings of the CHI Conference on Human Factors in Computing Systems},
    series    = {CHI '24},
    publisher = {Association for Computing Machinery},
    year      = {2024},
    doi       = {10.1145/3613904.3642587},
    url       = {https://dl.acm.org/doi/10.1145/3613904.3642587}
}

@inproceedings{Zoshak2021Beyond,
    author = {Zoshak, John and Dew, Kristin},
    title = {Beyond Kant and Bentham: How Ethical Theories are being used in Artificial Moral Agents},
    year = {2021},
    isbn = {9781450380966},
    publisher = {Association for Computing Machinery},
    address = {New York, NY, USA},
    url = {https://doi.org/10.1145/3411764.3445102},
    doi = {10.1145/3411764.3445102},
    booktitle = {Proceedings of the 2021 CHI Conference on Human Factors in Computing Systems},
    articleno = {590},
    numpages = {15},
    location = {Yokohama, Japan},
    series = {CHI '21}
}

@inproceedings{Khan2022Ethics,
    author    = {A. A. Khan and S. Badshah and P. Liang and M. Waseem and S. A. Khan and M. Shoaib and Y. Cui and S. Luo},
    title     = {Ethics of AI: A systematic literature review of principles and challenges},
    booktitle = {Proceedings of the 26th International Conference on Evaluation and Assessment in Software Engineering},
    year      = {2022},
    publisher = {Association for Computing Machinery},
    doi       = {10.1145/3530019.3530032},
    url       = {https://dl.acm.org/doi/10.1145/3530019.3530032}
}

@inproceedings{Henriques2025Feminist,
    author = {Henriques, Ana O and Carter, Anna R. L. and Severes, Beatriz and Talhouk, Reem and Strohmayer, Angelika and Pires, Ana Cristina and Gray, Colin M. and Montague, Kyle and Nicolau, Hugo},
    title = {A Feminist Care Ethics Toolkit for Community-Based Design: Bridging Theory and Practice},
    year = {2025},
    isbn = {9798400713941},
    publisher = {Association for Computing Machinery},
    address = {New York, NY, USA},
    url = {https://doi.org/10.1145/3706598.3713950},
    doi = {10.1145/3706598.3713950},
    booktitle = {Proceedings of the 2025 CHI Conference on Human Factors in Computing Systems},
    articleno = {396},
    numpages = {26},
    series = {CHI '25}
}

@article{Lam2025Confucian,
    title = {Building ethical virtual classrooms: Confucian perspectives on avatars and VR},
    journal = {Computers \& Education: X Reality},
    volume = {6},
    pages = {100092},
    year = {2025},
    issn = {2949-6780},
    doi = {https://doi.org/10.1016/j.cexr.2024.100092},
    url = {https://www.sciencedirect.com/science/article/pii/S2949678024000424},
    author = {Chi-Ming Lam},
}

@inproceedings{Kohno2023Ethical,
    author = {Kohno, Tadayoshi and Acar, Yasemin and Loh, Wulf},
    title = {Ethical frameworks and computer security trolley problems: foundations for conversations},
    year = {2023},
    isbn = {978-1-939133-37-3},
    publisher = {USENIX Association},
    address = {USA},
    booktitle = {Proceedings of the 32nd USENIX Conference on Security Symposium},
    articleno = {288},
    numpages = {18},
    location = {Anaheim, CA, USA},
    series = {SEC '23}
}

@book{Friedman2019Value,
    author    = {Batya Friedman and David G. Hendry},
    title     = {Value Sensitive Design: Shaping Technology with Moral Imagination},
    year      = {2019},
    publisher = {The MIT Press},
    address   = {Cambridge, MA},
    url       = {https://direct.mit.edu/books/monograph/4328/Value-Sensitive-DesignShaping-Technology-with}
}

@book{Dunne2013Speculative,
    author    = {Anthony Dunne and Fiona Raby},
    title     = {Speculative Everything: Design, Fiction, and Social Dreaming},
    year      = {2013},
    publisher = {The MIT Press},
    address   = {Cambridge, MA},
    isbn      = {9780262019842},
    url       = {https://mitpress.mit.edu/9780262019842/speculative-everything/}
}

@inproceedings{Wang2024Farsight,
    author = {Wang, Zijie J. and Kulkarni, Chinmay and Wilcox, Lauren and Terry, Michael and Madaio, Michael},
    title = {Farsight: Fostering Responsible AI Awareness During AI Application Prototyping},
    year = {2024},
    isbn = {9798400703300},
    publisher = {Association for Computing Machinery},
    url = {https://doi.org/10.1145/3613904.3642335},
    doi = {10.1145/3613904.3642335},
    booktitle = {Proceedings of the 2024 CHI Conference on Human Factors in Computing Systems},
    articleno = {976},
    numpages = {40},
    location = {Honolulu, HI, USA},
    series = {CHI '24}
}

@inproceedings{Bardzell2013Critical,
    author    = {Jeffrey Bardzell},
    title     = {What is "critical" about critical design?},
    booktitle = {Proceedings of the SIGCHI Conference on Human Factors in Computing Systems},
    year      = {2013},
    publisher = {Association for Computing Machinery},
    doi       = {10.1145/2470654.2466451},
    url       = {https://dl.acm.org/doi/10.1145/2470654.2466451}
}

@inproceedings{Vilaza2022Scoping,
    author = {Nunes Vilaza, Giovanna and Doherty, Kevin and McCashin, Darragh and Coyle, David and Bardram, Jakob and Barry, Marguerite},
    title = {A Scoping Review of Ethics Across SIGCHI},
    year = {2022},
    isbn = {9781450393584},
    publisher = {Association for Computing Machinery},
    address = {New York, NY, USA},
    url = {https://doi.org/10.1145/3532106.3533511},
    doi = {10.1145/3532106.3533511},
    booktitle = {Proceedings of the 2022 ACM Designing Interactive Systems Conference},
    pages = {137–154},
    numpages = {18},
    location = {Virtual Event, Australia},
    series = {DIS '22}
}

@inproceedings{Fiesler2020Teach,
    author = {Fiesler, Casey and Garrett, Natalie and Beard, Nathan},
    title = {What Do We Teach When We Teach Tech Ethics? A Syllabi Analysis},
    year = {2020},
    isbn = {9781450367936},
    publisher = {Association for Computing Machinery},
    address = {New York, NY, USA},
    url = {https://doi.org/10.1145/3328778.3366825},
    doi = {10.1145/3328778.3366825},
    booktitle = {Proceedings of the 51st ACM Technical Symposium on Computer Science Education},
    pages = {289–295},
    numpages = {7},
    location = {Portland, OR, USA},
    series = {SIGCSE '20}
}

@article{schrier2017designing,
  title={Designing role-playing video games for ethical thinking},
  author={Schrier, Karen},
  journal={Educational Technology Research and Development},
  volume={65},
  number={4},
  pages={831--868},
  year={2017},
  publisher={Springer}
}

@article{Brown2024Teaching,
    author    = {Noelle Brown and Benjamin Xie and Ella Sarder and Casey Fiesler and Natalie Garrett and Nathan Beard},
    title     = {Teaching Ethics in Computing: A Systematic Literature Review of ACM Computer Science Education Publications},
    journal   = {ACM Transactions on Computing Education},
    volume    = {24},
    number    = {1},
    pages     = {1--36},
    year      = {2024},
    doi       = {10.1145/3634685},
    url       = {https://dl.acm.org/doi/10.1145/3634685}
}

@article{Grosz2019Embedded,
    author    = {Barbara J. Grosz and David Gray Grant and Kate Vredenburgh and Jeff Behrends and Lily Hu and Alison Simmons and Jim Waldo},
    title     = {Embedded EthiCS: Integrating Ethics Across CS Education},
    journal   = {Communications of the ACM},
    volume    = {62},
    number    = {8},
    pages     = {54--61},
    year      = {2019},
    doi       = {10.1145/3330794},
    url       = {https://dl.acm.org/doi/10.1145/3330794}
}

@inproceedings{Ballard2019Judgment,
    author = {Ballard, Stephanie and Chappell, Karen M. and Kennedy, Kristen},
    title = {Judgment Call the Game: Using Value Sensitive Design and Design Fiction to Surface Ethical Concerns Related to Technology},
    year = {2019},
    isbn = {9781450358507},
    publisher = {Association for Computing Machinery},
    booktitle = {DIS '19: Proceedings of the 2019 on Designing Interactive Systems Conference},
    address = {New York, NY, USA},
    url = {https://doi.org/10.1145/3322276.3323697},
    doi = {10.1145/3322276.3323697},
    pages = {421–433},
    numpages = {13},
    location = {San Diego, CA, USA},
    series = {DIS '19}
}

@inproceedings{Jarzemsky2023Applies,
  title={" This Applies to the Real World": Student Perspectives on Integrating Ethics into a Computer Science Assignment},
  author={Jarzemsky, Julie and Paup, Joshua and Fiesler, Casey},
  booktitle={Proceedings of the 54th ACM Technical Symposium on Computer Science Education V. 1},
  pages={374--380},
  year={2023}
}

@inproceedings{Brown2022Shortest,
    author    = {Noelle Brown and Katherine South and Ella Sarder Wiese},
    title     = {The Shortest Path to Ethics in AI: An Integrated Assignment Where Human Concerns Guide Technical Decisions},
    booktitle = {ACM International Computing Education Research Conference (ICER 2022)},
    year      = {2022},
    pages     = {1--11},
    doi       = {10.1145/3477195.3532099},
    url       = {https://dl.acm.org/doi/10.1145/3477195.3532099}
}

@misc{katz2024thematic,
    title={Thematic Analysis with Open-Source Generative AI and Machine Learning: A New Method for Inductive Qualitative Codebook Development}, 
    author={Andrew Katz and Gabriella Coloyan Fleming and Joyce Main},
    year={2024},
    eprint={2410.03721},
    archivePrefix={arXiv},
    primaryClass={cs.CL},
    url={https://arxiv.org/abs/2410.03721}, 
}

@misc{Goldman2018Playbook,
    author    = {Ilana Lipsett and Lane Becker and Jake Dunagan},
    title     = {A Playbook for Ethical Technology Governance},
    year      = {2018},
    institution = {Institute for the Future},
    editor    = {Mark Frauenfelder},
    url       = {https://www.iftf.org/projects/a-playbook-for-ethical-tech-governance/},
    note      = {SR-2150}
}

@inproceedings{Klassen2022Run,
    author = {Klassen, Shamika and Fiesler, Casey},
    title = {"Run Wild a Little With Your Imagination": Ethical Speculation in Computing Education with Black Mirror},
    year = {2022},
    isbn = {9781450390705},
    publisher = {Association for Computing Machinery},
    address = {New York, NY, USA},
    url = {https://doi.org/10.1145/3478431.3499308},
    doi = {10.1145/3478431.3499308},
    booktitle = {Proceedings of the 53rd ACM Technical Symposium on Computer Science Education - Volume 1},
    pages = {836–842},
    numpages = {7},
    location = {Providence, RI, USA},
    series = {SIGCSE 2022}
}

@inproceedings{Luria2022Letters,
    author = {Luria, Michal and Candy, Stuart},
    title = {Letters from the Future: Exploring Ethical Dilemmas in the Design of Social Agents},
    year = {2022},
    isbn = {9781450391573},
    publisher = {Association for Computing Machinery},
    address = {New York, NY, USA},
    url = {https://doi.org/10.1145/3491102.3517536},
    doi = {10.1145/3491102.3517536},
    booktitle = {Proceedings of the 2022 CHI Conference on Human Factors in Computing Systems},
    articleno = {419},
    numpages = {13},
    location = {New Orleans, LA, USA},
    series = {CHI '22}
}

@inproceedings{Shapiro2021RolePlay,
    author = {Shapiro, Ben Rydal and Lovegall, Emma and Meng, Amanda and Borenstein, Jason and Zegura, Ellen},
    title = {Using Role-Play to Scale the Integration of Ethics Across the Computer Science Curriculum},
    year = {2021},
    isbn = {9781450380621},
    publisher = {Association for Computing Machinery},
    address = {New York, NY, USA},
    url = {https://doi.org/10.1145/3408877.3432525},
    doi = {10.1145/3408877.3432525},
    booktitle = {Proceedings of the 52nd ACM Technical Symposium on Computer Science Education},
    pages = {1034–1040},
    numpages = {7},
    location = {Virtual Event, USA},
    series = {SIGCSE '21}
}

@inproceedings{DeBoer2019DecidArch,
    author={de Boer, Remco C. and Lago, Patricia and Verdecchia, Roberto and Kruchten, Philippe},
    booktitle={2019 IEEE International Conference on Software Architecture Companion (ICSA-C)}, 
    title={DecidArch V2: An Improved Game to Teach Architecture Design Decision Making}, 
    year={2019},
    volume={},
    number={},
    pages={153-157},
    doi={10.1109/ICSA-C.2019.00034}
}

@inproceedings{Alidoosti2023Ethics,
    author={Alidoosti, Razieh and Lago, Patricia and Poort, Eltjo and Razavian, Maryam},
    booktitle={2023 IEEE 20th International Conference on Software Architecture Companion (ICSA-C)}, 
    title={Ethics-Aware DecidArch Game: Designing a Game to Reflect on Ethical Considerations in Software Architecture Design Decision Making}, 
    Organization={IEEE},
    year={2023},
    volume={},
    number={},
    pages={96-100},
    doi={10.1109/ICSA-C57050.2023.00031}
}

@article{So2024Dialogue,
    author    = {Hyo-Jeong So and Sung-Eun Kim},
    title     = {Dialogue Game-Based Learning for AI Ethics Education},
    journal   = {International Conference on Computers in Education (ICCE)},
    year      = {2024},
    url       = {https://library.apsce.net/index.php/ICCE/article/view/4894}
}

@incollection{Schrier2014Designing,
    author    = {Karen Schrier},
    title     = {Designing and using games to teach ethics and ethical thinking},
    booktitle = {Learning, Education and Games: Volume One: Curricular and Design Considerations},
    editor    = {Karen Schrier},
    pages     = {73--91},
    publisher = {ETC Press},
    year      = {2014},
    url       = {https://press.etc.cmu.edu/books/learning-education-games/1/1}
}

@article{Schrier2015EPIC,
    author    = {Karen Schrier},
    title     = {EPIC: A framework for using video games in ethics education},
    journal   = {Journal of Moral Education},
    volume    = {44},
    number    = {4},
    pages     = {393--424},
    year      = {2015},
    doi       = {10.1080/03057240.2015.1095168},
    url       = {https://www.tandfonline.com/doi/full/10.1080/03057240.2015.1095168}
}

@inproceedings{Castro2023Piloting,
    author = {Castro, Francisco and Raipura, Sahitya and Conboy, Heather and Haas, Peter and Osterweil, Leon and Arroyo, Ivon},
    title = {Piloting an Interactive Ethics and Responsible Computing Learning Environment in Undergraduate CS Courses},
    year = {2023},
    isbn = {9781450394314},
    publisher = {Association for Computing Machinery},
    address = {New York, NY, USA},
    url = {https://doi.org/10.1145/3545945.3569753},
    doi = {10.1145/3545945.3569753},
    booktitle = {Proceedings of the 54th ACM Technical Symposium on Computer Science Education V. 1},
    pages = {659–665},
    numpages = {7},
    location = {Toronto ON, Canada},
    series = {SIGCSE 2023}
}

@article{Wu2025AILEGO,
    author    = {Muzhe Wu and Yanzhi Zhao and Shuyi Han and Michael Xieyang Liu and Hong Shen},
    title     = {AI LEGO: Scaffolding Cross-Functional Collaboration in Industrial Responsible AI Practices during Early Design Stages},
    journal   = {arXiv preprint arXiv:2505.10300},
    year      = {2025},
    url       = {https://arxiv.org/abs/2505.10300},
    doi       = {10.48550/arXiv.2505.10300}
}

@inproceedings{ElsayedAli2023Responsible,
    author = {Elsayed-Ali, Salma and Berger, Sara E and Santana, Vagner Figueredo De and Becerra Sandoval, Juana Catalina},
    title = {Responsible \& Inclusive Cards: An Online Card Tool to Promote Critical Reflection in Technology Industry Work Practices},
    year = {2023},
    isbn = {9781450394215},
    publisher = {Association for Computing Machinery},
    address = {New York, NY, USA},
    url = {https://doi.org/10.1145/3544548.3580771},
    doi = {10.1145/3544548.3580771},
    booktitle = {Proceedings of the 2023 CHI Conference on Human Factors in Computing Systems},
    articleno = {5},
    numpages = {14},
    location = {Hamburg, Germany},
    series = {CHI '23}
}

@article{Bucinca2023AHA,
    author    = {Zana Bu{\c{c}}inca and Chau Minh Pham and Maurice Jakesch and Marco Tulio Ribeiro and Alexandra Olteanu and Saleema Amershi},
    title     = {AHA!: Facilitating AI Impact Assessment by Generating Examples of Harms},
    journal   = {arXiv preprint arXiv:2306.03280},
    year      = {2023},
    url       = {https://arxiv.org/abs/2306.03280},
    doi       = {10.48550/arXiv.2306.03280}
}

@article{Feffer2023AIID,
    author    = {Michael Feffer and Nikolas Martelaro and Hoda Heidari},
    title     = {The AI Incident Database as an Educational Tool to Raise Awareness of AI Harms: A Classroom Exploration of Efficacy, Limitations, \& Future Improvements},
    journal   = {arXiv preprint arXiv:2310.06269},
    year      = {2023},
    url       = {https://arxiv.org/abs/2310.06269},
    doi       = {10.48550/arXiv.2310.06269}
}

@article{Yu2024Who,
    author = {Ruoyu, Yu and Feng, Jingdan and Wang, Kai and Yang, Luoyao and Feng, Jiao},
    year = {2024},
    month = {07},
    pages = {1090-1106},
    title = {Who are you talking to? How chatbot identity disclosure affects service satisfaction},
    volume = {41},
    journal = {Journal of Travel \& Tourism Marketing},
    doi = {10.1080/10548408.2024.2369755}
}

@inproceedings{Mozafari2020Chatbot,
    author = {Meyer, Nika and Weiger, Welf and Hammerschmidt, Maik},
    year = {2020},
    month = {09},
    booktitle = {Proceedings of the International Conference on Information Systems (ICIS)},
    title = {The Chatbot Disclosure Dilemma: Desirable and Undesirable Effects of Disclosing the Non-Human Identity of Chatbots}
}

@inproceedings{Mozafari2021Resolving,
  title={Resolving the chatbot disclosure dilemma: leveraging selective self-presentation to mitigate the negative effect of chatbot disclosure},
  author={Mozafari, Nika and Weiger, Welf H and Hammerschmidt, Maik},
  booktitle = {Proceedings of the 54th Hawaii International Conference on System Sciences},
  year={2021}
}

@inproceedings{Lee2025Conversational,
    author = {Lee, Soohwan and Hwang, Seoyeong and Kim, Dajung and Lee, Kyungho},
    title = {Conversational Agents as Catalysts for Critical Thinking: Challenging Social Influence in Group Decision-making},
    year = {2025},
    isbn = {9798400713958},
    publisher = {Association for Computing Machinery},
    address = {New York, NY, USA},
    url = {https://doi.org/10.1145/3706599.3719792},
    doi = {10.1145/3706599.3719792},
    booktitle = {Proceedings of the Extended Abstracts of the CHI Conference on Human Factors in Computing Systems},
    articleno = {154},
    numpages = {12},
    series = {CHI EA '25}
}

@article{May2014Influence,
    title={The influence of business ethics education on moral efficacy, moral meaningfulness, and moral courage: A quasi-experimental study},
    author={May, Douglas R and Luth, Matthew T and Schwoerer, Catherine E},
    journal={Journal of Business Ethics},
    volume={124},
    number={1},
    pages={67--80},
    year={2014},
    publisher={Springer}
}

@article{Wen2025Trust,
    title={Trust and AI weight: human-AI collaboration in organizational management decision-making},
    author={Yanjun Wen and Jiale Wang and Xiaoxi Chen},
    journal={Frontiers in Organizational Psychology},
    year={2025},
    url={https://api.semanticscholar.org/CorpusID:279478875}
}

@misc{Song2024Multi,
      title={Multi-Agents are Social Groups: Investigating Social Influence of Multiple Agents in Human-Agent Interactions}, 
      author={Tianqi Song and Yugin Tan and Zicheng Zhu and Yibin Feng and Yi-Chieh Lee},
      year={2024},
      eprint={2411.04578},
      archivePrefix={arXiv},
      primaryClass={cs.AI},
      url={https://arxiv.org/abs/2411.04578}, 
}

@inproceedings{Tran2024Student,
    author = {Tran, Michelle and Fiesler, Casey},
    title = {"It's Not Exactly Meant to Be Realistic": Student Perspectives on the Role of Ethics In Computing Group Projects},
    year = {2024},
    isbn = {9798400704758},
    publisher = {Association for Computing Machinery},
    address = {New York, NY, USA},
    url = {https://doi.org/10.1145/3632620.3671109},
    doi = {10.1145/3632620.3671109},    
    booktitle = {Proceedings of the 2024 ACM Conference on International Computing Education Research - Volume 1},
    pages = {517–526},
    numpages = {10},
    location = {Melbourne, VIC, Australia},
    series = {ICER '24}
}

@incollection{Tirri2011ESS,
    author    = {Kirsi Tirri and Petri Nokelainen},
    title     = {Ethical Sensitivity Scale},
    booktitle = {Measuring Multiple Intelligences and Moral Sensitivities in Education},
    editor    = {Kirsi Tirri and Petri Nokelainen},
    pages     = {59--75},
    series    = {Moral Development and Citizenship Education},
    volume    = {5},
    publisher = {SensePublishers},
    address   = {Rotterdam},
    year      = {2011},
    doi       = {10.1007/978-94-6091-758-5_4},
    url       = {https://doi.org/10.1007/978-94-6091-758-5_4}
}

@article{kreismann2021business,
    title={Business ethics training in human resource development: A literature review},
    author={Kreismann, Dominic and Talaulicar, Till},
    journal={Human Resource Development Review},
    volume={20},
    number={1},
    pages={68--105},
    year={2021},
    publisher={Sage Publications Sage CA: Los Angeles, CA}
}

@article{senior2014learning,
    title={Learning in friendship groups: developing students’ conceptual understanding through social interaction},
    author={Senior, Carl and Howard, Chris},
    journal={Frontiers in psychology},
    volume={5},
    pages={1031},
    year={2014},
    publisher={Frontiers Media SA}
}

@article{Jasemi2022,
  author  = {Jasemi, Madineh and others},
  title   = {Educating ethics codes by lecture or role-play; which one improves nursing students' ethical sensitivity and ethical performance more? A quasi-experimental study},
  journal = {Journal of Professional Nursing},
  year    = {2022},
  volume  = {40},
  pages   = {122--129}
}

@article{Pourghaznein2015,
  author       = {Pourghaznein, T. and Sabeghi, H. and Shariatinejad, K.},
  title        = {Effects of e-learning, lectures, and role playing on nursing students' knowledge acquisition, retention and satisfaction},
  journal      = {Medical Journal of the Islamic Republic of Iran},
  year         = {2015},
  volume       = {29},
  pages        = {162},
  date         = {2015-01-25},
  pmid         = {26000257},
  pmcid        = {PMC4431360}
}

@article{baker1999facilitator,
  title={The Facilitator's Guide to Participatory Decision-Making},
  author={Baker, Lynda Lieberman},
  journal={Group Facilitation},
  number={1},
  pages={52},
  year={1999},
  publisher={International Association of Facilitators}
}

@book{kaner2014facilitator,
  title={Facilitator's guide to participatory decision-making},
  author={Kaner, Sam},
  year={2014},
  publisher={John Wiley \& Sons}
}

@article{Byun2018Performative,
    author    = {Young Seok Byun and Sung Kyu Park and Joon Sakong and Man Joong Jeon},
    title     = {Performance assessment on the Korean Computerized Neurobehavioral Test using a mobile device and a conventional computer: an experimental study},
    journal   = {Environmental Health and Toxicology},
    volume    = {33},
    pages     = {e2018017},
    year      = {2018},
    doi       = {10.5620/eht.e2018017},
    url       = {https://www.ncbi.nlm.nih.gov/pmc/articles/PMC6114805/}
}

@article{Kim2016Korean,
    author    = {Dae Yeon Kim and Yong Seok Han and Sin Young Yeh and Youn Chul Chung},
    title     = {Validity of Korean Version Reading Speed Application and Measurement of Reading Speed: Pilot Study},
    journal   = {Journal of Korean Ophthalmological Society},
    volume    = {57},
    number    = {4},
    pages     = {642--649},
    year      = {2016},
    url       = {https://kci.go.kr/kciportal/ci/sereArticleSearch/ciSereArtiView.kci?sereArticleSearchBean.artiId=ART002099342}
}

@article{lenskjold2023should,
  title={Should artificial intelligence have lower acceptable error rates than humans?},
  author={Lenskjold, Anders and Nybing, Janus Uhd and Trampedach, Charlotte and Galsgaard, Astrid and Brejneb{\o}l, Mathias Willadsen and Raaschou, Henriette and Rose, Martin H{\o}yer and Boesen, Mikael},
  journal={BJR| Open},
  volume={5},
  number={1},
  pages={20220053},
  year={2023},
  publisher={Oxford University Press}
}

@article{cai2025understanding,
  title={Understanding consumer reactions to chatbot service failures: Evidence from a Wizard-of-Oz experiment},
  author={Cai, Na and Heo, Jeakang and Yan, Jinzhe},
  journal={Acta Psychologica},
  volume={253},
  pages={104707},
  year={2025},
  publisher={Elsevier}
}

@article{zhang2013structural,
  title={The structural features and the deliberative quality of online discussions},
  author={Zhang, Weiyu and Cao, Xiaoxia and Tran, Minh Ngoc},
  journal={Telematics and informatics},
  volume={30},
  number={2},
  pages={74--86},
  year={2013},
  publisher={Elsevier}
}

@article{mozafari2022trust,
  title={Trust me, I'm a bot--repercussions of chatbot disclosure in different service frontline settings},
  author={Mozafari, Nika and Weiger, Welf H and Hammerschmidt, Maik},
  journal={Journal of Service Management},
  volume={33},
  number={2},
  pages={221--245},
  year={2022},
  publisher={Emerald Publishing Limited}
}

@inproceedings{bo2024disclosures,
  title={Disclosures \& disclaimers: Investigating the impact of transparency disclosures and reliability disclaimers on learner-LLM interactions},
  author={Bo, Jessica Y and Kumar, Harsh and Liut, Michael and Anderson, Ashton},
  booktitle={Proceedings of the AAAI Conference on Human Computation and Crowdsourcing},
  volume={12},
  pages={23--32},
  year={2024}
}

@article{chan2023students,
  title={Students’ voices on generative AI: Perceptions, benefits, and challenges in higher education},
  author={Chan, Cecilia Ka Yuk and Hu, Wenjie},
  journal={International Journal of Educational Technology in Higher Education},
  volume={20},
  number={1},
  pages={43},
  year={2023},
  publisher={Springer}
}

@inproceedings{ha2024clochat,
  title={CloChat: Understanding how people customize, interact, and experience personas in large language models},
  author={Ha, Juhye and Jeon, Hyeon and Han, Daeun and Seo, Jinwook and Oh, Changhoon},
  booktitle={Proceedings of the 2024 CHI Conference on Human Factors in Computing Systems},
  pages={1--24},
  year={2024}
}

@inproceedings{park2023generative,
  title={Generative agents: Interactive simulacra of human behavior},
  author={Park, Joon Sung and O'Brien, Joseph and Cai, Carrie Jun and Morris, Meredith Ringel and Liang, Percy and Bernstein, Michael S},
  booktitle={Proceedings of the 36th annual acm symposium on user interface software and technology},
  pages={1--22},
  year={2023}
}

@misc{simmons2018sample,
  title        = {Sample Rubrics for Discussion Forums},
  author       = {{Blended Learning \& Academic Technology, Simmons College}},
  year         = {2018},
  month        = aug,
  howpublished = {PDF document, Simmons College},
  note         = {Available at \url{https://effectiveness.syr.edu/wp-content/uploads/2018/08/sample_rubrics_for_discussions_Simmons.pdf}},
  institution  = {Simmons College}
}

@misc{OpenAI2024GPT4o,
  title        = {Hello GPT-4o},
  author       = {{OpenAI}},
  year         = {2024},
  month        = may,
  howpublished = {OpenAI Blog},
  note         = {Available at \url{https://openai.com/ko-KR/index/hello-gpt-4o/}, Accessed: 2026-07-31}
}

@article{Zouaghi2025Temperature,
  title     = {Balancing Performance and Innovation in AI-Driven Supply Chains Through Temperature-Scaled Hallucination Control},
  author    = {Zouaghi, Iskander and Beldjoudi, Samia and Gunasekaran, Angappa and Sehrane, Selim},
  journal   = {International Journal of Production Research},
  year      = {2025},
  publisher = {Taylor \& Francis},
  doi       = {10.1080/00207543.2025.2601264},
  note      = {Advance online publication}
}

@article{Wang2024LLMAgentSurvey,
  title     = {A Survey on Large Language Model Based Autonomous Agents},
  author    = {Wang, Lei and Ma, Chen and Feng, Xueyang and Zhang, Zeyu and Yang, Hao and Zhang, Jingsen and Chen, Zhiyuan and Tang, Jiakai and Chen, Xu and Lin, Yankai and Zhao, Wayne Xin and Wei, Zhewei and Wen, Ji-Rong},
  journal   = {Frontiers of Computer Science},
  volume    = {18},
  number    = {6},
  pages     = {186345},
  year      = {2024},
  publisher = {Springer},
  doi       = {10.1007/s11704-024-40231-1},
  note      = {arXiv:2308.11432}
}

\lstdefinestyle{code}{
  basicstyle=\ttfamily\small,
  numberstyle=\tiny,
  stepnumber=1,
  numbersep=8pt,
  showstringspaces=false,
  breaklines=true,
  breakatwhitespace=true,
  frame=single,
  tabsize=2,
  captionpos=b
}
\clearpage
\appendix
\section{Appendix: Measurement Instruments and Peer Rating Analysis}
\label{sec:appen-a}
This appendix presents the full item lists of the two measurement instruments used in the study: the Ethical Sensitivity in Technology Design scale and the Peer Rating measure, as shown in Tables~\ref{tab:Ethical sensitivity} and \ref{tab:Peer Rating}. And the detailed statistical results of peer rating are presented in Table \ref{tab:peer_rating_analysis}.

\begin{table}[H]
\centering
\caption{Ethical Sensitivity in Technology Design}
\label{tab:Ethical sensitivity}
\small
\begin{tabular}{p{0.95\linewidth}}
\toprule
\textbf{Ethical Issue Awareness} \\
1. I am aware that technology design choices can affect different stakeholders in different ways. \\
2. I understand that product design can unintentionally disadvantage certain groups. \\
3. I recognize that product and AI design often raise ethical issues. \\
4. I am aware that there are usually multiple perspectives on how to handle ethical challenges. \\ %

\textbf{Ethical Consequence Reasoning} \\
5. I can identify situations where bias might shape technology outcomes. \\
6. I can identify potential ethical problems that could arise in technology projects. \\
7. I can identify possible downstream consequences of technology decisions. \\
8. I can identify different alternatives when faced with an ethically problematic situation. \\
\bottomrule
\end{tabular}
\end{table}

\begin{table}[H]
\centering
\caption{Peer Rating Dimensions and Items}
\label{tab:Peer Rating}
\small
\begin{tabularx}{\linewidth}{lX}
\hline
\textbf{Dimension} & \textbf{Items} \\
\hline
General Contribution&
Contributed meaningfully to group discussions.   \newline
Demonstrated a cooperative and supportive attitude.  \newline
Contributed significantly to the success of the discussion.    \newline
Actively engaged in the discussion.   \\ \hline

Diversity&
Contributed to broadening the scope of thinking with creative ideas.   \newline
Offered creative or unexpected ideas during the discussion.   \\ \hline

Influence&
I think the perspective of this participant was important.   \newline
I was influenced by the opinion of this participant.   \\ \hline

\hline
\end{tabularx}
\end{table}
\begin{table}[H]
\centering
\caption{ Kruskal--Wallis test results by group (non-parametric, selected based on distributional assumption checks; see Section 4.4). Values are mean (SD).  Dunn's post-hoc comparisons (Holm-corrected) are reported against human peers; no significant differences were observed among the AI agents (all n.s.). A parallel one-way ANOVA is reported alongside for comparability with prior work.}
\label{tab:peer_rating_analysis}
\small
\begin{tabular}{p{3.3cm} p{3.6cm} p{6.8cm}}
\hline
\textbf{Dimension} & \textbf{ Kruskal--Wallis} & \textbf{Mean (SD) and Post-hoc Results} \\
\hline
General contribution &
$H(3)=21.97$ \newline
$p<.001$ \newline
$\varepsilon^2=.086$ \newline
(ANOVA: $F(3,221)=5.99$, \newline
$p<.001$, $\eta^2=.075$) &
Human: $M=3.31$ ($SD=0.63$) \newline
Care: $M=2.99$ ($SD=0.45$),  $p=.002$ \newline
Deontology : $M=3.03$ ($SD=0.59$),  $p=.012$ \newline
Pragmatic: $M=2.91$ ($SD=0.66$),  $p<.001$ \\
\hline
Diversity &
$H(3)=35.79$ \newline
$p<.001$ \newline
$\varepsilon^2=.148$ \newline
(ANOVA: $F(3,221)=14.18$, \newline
$p<.001$, $\eta^2=.161$) &
Human: $M=3.06$ ($SD=0.80$) \newline
Care: $M=2.28$ ($SD=0.77$), $p<.001$ \newline
Deontology: $M=2.34$ ($SD=0.81$), $p<.001$ \newline
Pragmatic: $M=2.39$ ($SD=0.90$), $p<.001$ \\
\hline
Influence &
$H(3)=19.38$ \newline
$p<.001$ \newline
$\varepsilon^2=.074$ \newline
(ANOVA: $F(3,221)=6.56$, \newline
$p<.001$, $\eta^2=.082$) &
Human: $M=3.18$ ($SD=0.67$) \newline
Care: $M=2.78$ ($SD=0.68$),  $p=.007$ \newline
Deontology: $M=2.74$ ($SD=0.77$),  $p=.006$ \newline
Pragmatic: $M=2.71$ ($SD=0.82$),  $p=.003$ \\
\hline
\end{tabular}
\end{table}

\clearpage

\section{Appendix: Program Structure}
\begin{figure}[H]
    \centering
    \includegraphics[width=\linewidth]{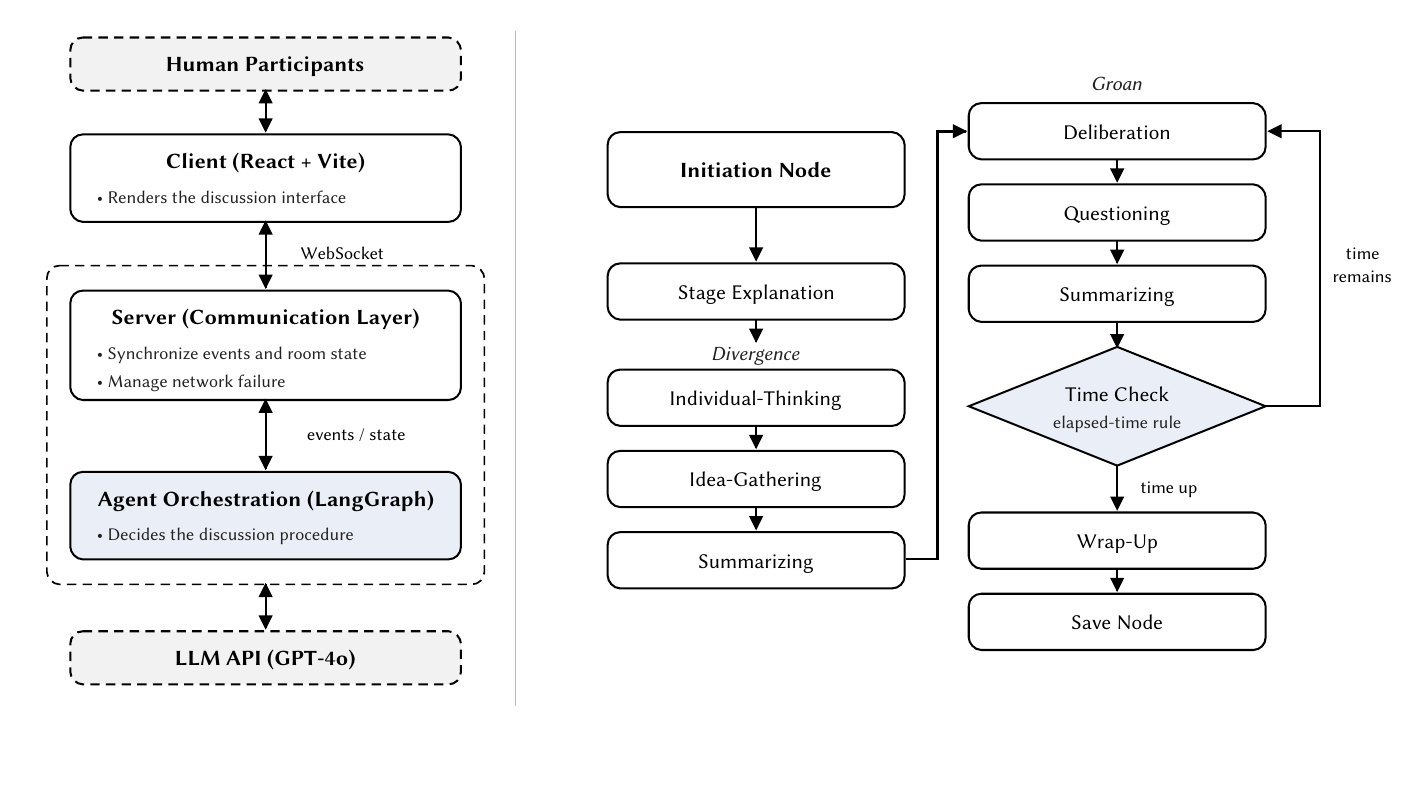}\caption{Overall architecture of the system (left) and node structure of agent orchestration module (right).}
    \Description{A two-panel figure. Panel (a) shows the client, server, and LangGraph agent orchestration modules connected in a vertical stack, with human participants above the client and the LLM API below the orchestration module. Panel (b) shows the state graph, running from the initiation node through stage explanation, the divergence phases, the groan phases, a time check, and wrap-up, looping over the four stages before reaching a save node. The phases are laid out in two columns, with the Groan phases beginning the second column.}
    \label{fig:architecture}
\end{figure}
\label{sec:program_structure}
This appendix describes the implementation of our system. Prompts referenced by each node are listed in Appendix~\ref{sec:appen-b}. Our codes are available at: \url{https://anonymous.4open.science/r/EthicsTrainingAgents/}.
\subsection{Architecture Overview}
The system is organized into three modules (Fig.~\ref{fig:architecture}a). The client
renders the discussion interface and transmits participant actions. The server relays
events and ensures that all LLM agents and human participants share the same state. The
agent orchestration module, built on LangGraph, makes all decisions concerning the
discussion procedure.

\subsection{Server and Communication Layer}
We used WebSocket because the discussion requires participants to observe one another's actions in real time (e.g., raising a hand, a transfer of the speaking turn, and an incoming utterance).
\subsubsection{Event Broadcasting}
The server broadcasts two kinds of updates. Events are actions generated by a participant or an LLM agent, namely raising a hand, submitting an utterance, and passing the speaking turn. Room state is the current condition of the discussion, comprising the current stage, whether hand raising and utterance submission are currently permitted, the start times of the discussion and of the current stage, the participant list, and the summary set produced by the facilitator. Room state is rebroadcast in full whenever any of its components change, so that every client converges on the same view without reconstructing state from an event history.

For simplicity, system messages that address a single participant (e.g., informing a participant that the floor has been given to them) are broadcast to all clients and filtered client-side so that only the intended recipient renders them. These messages contain only turn-taking instructions and no discussion content, and participants interacted with the system through the provided interface.
\subsubsection{Fault Tolerance}
The server and the orchestration module run on the same physical machine, so that no
network issues were observed between the two modules. Participants, however, used laptops
connected to the Wi-Fi at the place where the experiment was held. Therefore, connection problems could occur depending
on the network conditions. We therefore designed the communication layer to tolerate
transient disconnection. Every message carries a unique identifier, and clients discard
messages whose identifier they have already processed, so that retransmission cannot
produce duplicate utterances. Also, when a socket reconnects, the server retransmits the full
discussion history and the current room state. Therefore, although the socket was lost in connection, the client can recover its entire
view after the network reconnects. There was no critical issue reported by the participant during the experiment.
\subsection{Client}
\label{subsec:client}

The client is implemented in React with Vite. Because a client connects to exactly one discussion room, the WebSocket connection is managed as a singleton pattern. The room
state broadcast by the server is held in React hooks, so that every component depending on it re-renders whenever an update arrives. The interface is therefore driven by the
broadcast room state rather than by local inference about the
discussion's progress: the client selects the interface configuration for the current
stage, and enables the utterance input, the hand-raise button, and the turn-pass button according to the permissions in the room state.
\subsection{Agent Orchestration}
\paragraph{Rationale for Choosing LangGraph}
The discussion procedure is hierarchical: four stages, each decomposed into phases, with each agent's response itself decomposed into a few generation steps. We implemented this procedure as an explicit state graph in LangGraph for two reasons. First, representing phases as nodes makes the transition rules explicit and easily inspectable. Second, LangGraph's development interface allows a subgraph of any granularity to be executed in isolation, which makes testing easier.
\paragraph{Graph Structure}
Figure~\ref{fig:architecture}b shows critical nodes implemented in the graph. The \emph{initiation} node loads the agent configurations, instantiates a virtual client for each LLM agent, and retrieves the list of human participants from the server. Each discussion phase is implemented as a corresponding node, and each node invokes the prompts listed in Appendix~\ref{sec:appen-b}. A time-check node applies the elapsed-time rule and routes the discussion either back to the Deliberation and Questioning cycle or forward to Wrap-up. Time check is held at the end of each phase, so an utterance in progress is never interrupted. A terminal node writes the complete discussion record to a JSON file. Exceptionally, the Write Reviews stage bypasses the phase structure and routes directly to review composition.
\paragraph{Event Observation and State Application}
Agent nodes must react to events that occur while the graph is executing, but reading these events directly during a node's execution would cause the discussion state to change midway through an LLM call. We therefore separated observation from application. An observer collects incoming events from the server continuously and stores them in a pending queue. A node that depends on the discussion state drains the queue at its beginning and applies the pending events to its state in one step; the state is then fixed for the remainder of the node's execution. This ensures that no event is dropped while a long-running API call is in flight, and that a node's state is stable for each node.

\subsubsection{Constrained Output Format}
We used the OpenAI Structured Outputs interface in three places to prevent LLM agents from referring to content that did not exist in the discussion. In dialogue generation, an LLM agent was required to emit the speaker and the utterance content together, so that an utterance could not be attributed to the wrong participant. In answerer selection, the set of human participants and LLM agents was defined as an enumerated type, and the facilitator selected one member of that set, which removes the possibility of designating someone who was not taking part in the discussion. In question generation, an LLM agent was required to state both the idea it was asking about and the participant who had proposed that idea, which prevents it from asking about an idea that had never been raised.

\subsubsection{State Visible to Agents}
An LLM agent is given the utterance history of the current stage together with the selected summaries from all preceding stages, which accumulate as stages progress. Reviews written during the Write Reviews stage enter later stages in the same summary format as other selected ideas. LLM agents do not maintain private memory; they refer only to this shared record. This defines the context supplied to each prompt in Appendix~\ref{sec:appen-b}.

\paragraph{Response Latency}
An agent utterance is released at the later of two times: When generation completes and when the pre- and post-utterance delay has elapsed. In internal testing, the response time of ChatGPT took between 800\,ms and 2{,}000\,ms, depending on the length of the response, whereas the delay for an utterance of ten words or more is at least 4{,}000\,ms. Because most agent utterances exceeded ten words, model response time had little effect on the interval participants observed, except when the API call failed several times, which happens very rarely. API call failures were handled by retry. And no session was aborted or visibly interrupted by an API failure.
\subsubsection{Experimenter Involvement}
The experimenter started each session and initiated stage transitions from the LangGraph development interface, and measured the two-minute individual reflection period at the initiation node to implement the self-thinking phase. The experimenter did not intervene after the self-thinking phase. All facilitation decisions were produced by the facilitator agent and the system.

\clearpage
\section{Appendix: Used Prompts}
\label{sec:appen-b}
Because the full prompt set is lengthy, the appendix reports only the core prompts that define the agent roles and control flow; the remaining supporting prompts follow the same structure and are omitted for brevity. For all prompts, explanations for each stage were adapted from the explanation of the original Judgment Call paper. Our prompts are also available at: \url{https://anonymous.4open.science/r/EthicsTrainingAgents/}, same as code.
\subsection{Common Prompt}
\subsubsection{Common Speech Rules}
The following "common speech rules" prompt is included by default in the prompts for all LLM agents' dialogue generating prompts. Some rules were slightly adapted to better fit each behavior (e.g., “Do not insert actions unrelated to the discussion content, like 'raise your hand' or 'put your hand down.'” was omitted from the stacking prompt of LLM facilitator).
\begin{lstlisting}[style=code]
    1. Do not include greetings such as "Hello."
    2. Do not insert actions unrelated to the discussion content, like "raise your hand" or "put your hand down."
    3. Do not say anything that is not related to your own opinion, like "If you have any other opinions, please speak up."
    4. Only talk about content related to the current discussion stage.
    5. Speak like a human.
    6. Write in a way that is easy for ordinary people to understand.
\end{lstlisting}
\subsubsection{Scenario Prompt}\hfill\break
\label{sec:scenario}
\begin{lstlisting}[style=code]
    Product Name : "Memora" - Memory Augmentation Smart Lens
    Product Description: Memora is a tiny, wearable smart contact lens that records memories based on the user's vision and enhances them using AI technology. It has a camera, storage, and wireless communication module built into the lens, automatically recording important moments in daily life and providing summary and restoration functions through AI.
    Main Features
        Eye Blink Control for Recording
            Users can control the recording start and stop by naturally blinking their eyes.
        Real-Time Video Streaming
            Users can stream their view in real-time to friends, family, and others to share the scene together.
        Automatic Summary and Highlight Generation
            AI automatically recognizes "important moments" in the user's daily life and automatically generates a summary and highlight video, then sends it to the user.
        Memory Assistance Function
            Users can call up specific memories (e.g., class or meeting content recorded visually) to restore them.
        Facial Recognition-Based Hint Generation
            When meeting someone familiar, AI provides a hint based on past memories about "where I've seen this person" to help identify them.
\end{lstlisting}
\subsection{Facilitator Prompt}
\subsubsection{Stage Explanation Prompt}\hfill\break
\begin{lstlisting}[style=code]
    You are a facilitator of an ethical discussion. Your task is to explain the stages and the scenario of the discussion in order.
    The stages of the discussion are as follows:
        {brief explanation for each stage}
    Current stage and discussion are as follows: 
        {current stage explanation}
        {current discussion scenario}
    [If previous stage exists]
        The ideas from the previous discussion stage {previous stage} are as follows:
        {summary of previous stage ideas}
    Explain the current discussion stage and scenario to the participants. Note that all content has been explained before, so please explain in 2 sentences or less.
\end{lstlisting}
\subsubsection{Stacking Prompt}\hfill\break
\begin{lstlisting}[style=code]
    You are the discussion facilitator for an ethics discussion. 
        Your role is to announce the idea-gathering time to the participants and guide them to think in line with the current discussion stage.
        
        {brief explanation for 5 stages}
        {current stage explanation}
        {current discussion scenario}
        
        Write a message that encourages participants to freely suggest ideas for {time limit}seconds.
        Clearly specify the time, for example: "For the next X minutes and Y seconds, please feel free to share your ideas."
        
        Follow these rules:
        1. Guide participants to think in accordance with the current discussion stage.
        2. For example, in the "Identify Stakeholders" stage, encourage them to think about "who might be affected by the product."
        3. The speaker's name should be written as "Facilitator."
        4. Keep the message concise, within 50 words or less.
        5. Participants already know the discussion stages. Do not re-explain the stages.
\end{lstlisting}
\subsubsection{Encourage Prompt}\hfill\break
\begin{lstlisting}[style=code]
    You are the discussion facilitator for an ethics discussion. 
        You are trying to encourage participants to share their opinions, but currently, no one has raised their hand.
        Write a message that strongly encourages active participation.
        Emphasize the importance of the discussion and how valuable each person's opinion is, 
        and guide them to share opinions related to the current discussion stage.
        
        {brief explanation for 5 stages}
        {current stage explanation}
        {current discussion scenario}
          
        During the discussion, please follow these rules:
            [common speech rule]
            1. Encourage participants to focus on the current discussion stage.
            2. If they begin talking about something outside the current stage, gently guide them back.
               For example, if the stage is "Identify Stakeholders" but they talk about product defects or design considerations, 
               encourage them by saying, "Please focus on content related to the current discussion stage."
            3. The speaker's name should be written as "Facilitator."
            4. Write your message within three sentences.
            5. Clearly mention at the beginning that no one has raised their hand yet.
\end{lstlisting}
\subsubsection{Summarize Prompt (Add Operation)}\hfill\break
\begin{lstlisting}[style=code]
    You are a dialogue summarization assistant. Your task is to analyze conversations and determine whether new summaries need to be added.
        {current stage explanation}
        {current discussion scenario}
    Please determine whether new summaries should be added that align with the purpose of the current discussion stage.
    You will be given discussion content. For the discussion content, you must perform the following:
    1. Check if there are any ideas in the discussion content that are not included in the summaries. Be careful not to miss even very briefly mentioned ideas (ideas expressed in just one or two words).
    2. If there is one or more ideas not included, create one operation for each idea, set the operationType to "add" and add the content.
    3. If all ideas are already included, create a single operation in a single operationList, set the operationType to "finish" and leave the content empty.
    
    Please proceed with summarization as follows:
    {summarization prompt for each stage}
    
    If the dialogue content is not related to the purpose of the current stage or is already appropriately summarized, set the operation to "finish" and leave the content empty.
    
    Current summarized idea list (do not add ideas already included here):
    {list of summary title/content}
    
    The dialogue content to analyze is as follows:
    {discussion between the last summary call and the current summary call}
    
    Please analyze this dialogue and determine whether new summaries need to be added.
\end{lstlisting}
\subsubsection{Summarize Prompt (Fix Operation)}\hfill\break
\begin{lstlisting}[style=code]
     You are a dialogue summarization assistant. Your task is to analyze conversations and determine whether existing summaries need to be modified.
        {current stage explanation}
        {current discussion scenario}
    Please determine whether existing summaries should be modified to align with the purpose of the current discussion stage.
    
    For the dialogue content, you must perform the following:
    1. Analyze the content to identify ideas or points.
    2. Check if the summary content matches the current content.
    3. If existing summaries need to be modified, create one operation for each summary that needs modification, set the operationType to "modify" and modify the content or classification.
    4. If there is no need to modify existing summaries, create a single operation in a single operationList, set the operationType to "finish" and leave the content empty.
    
    Please proceed with summarization as follows:
    {summarization prompt for each stage}
    
    If the dialogue content is not related to the purpose of the current stage or is already appropriately summarized, set the operation to "finish" and leave the content empty.
    
    The dialogue content to analyze is as follows:
    {discussion between the last summary call and the current summary call}
    
    Current summarized content:
    {list of summarized item}
    
    Please analyze this dialogue and determine whether existing summaries need to be modified.
\end{lstlisting}
\subsubsection{Summarize Prompt (Stakeholder Stage)}\hfill\break
Since we enabled Structured Outputs when using the OpenAI API, the LLM consistently adhered to the required output format and did not produce any format-violating outputs. Below is an example for the stakeholder stage. Note that the types were defined specifically for each stage. In the stakeholder stage, the model was required to classify each item as one of direct, indirect, or excluded stakeholders. In all other stages, the model selected the most relevant idea from the previous stage’s outputs, allowing the classification types to remain flexible depending on the discussion content. For example, if a student proposed the idea that ``Face detection could be privacy invasive'' at the problem stage after examining a passerby's review, it was classified under the type "Review from passerby."
\begin{lstlisting}[style=code]
Identify the stakeholder proposed by the user and summarize it. You should summarize by following the structure.
    title : Name of the stakeholder proposed by the user.
    content : Content of how the stakeholder proposed by the user is related to the product. (If the user did not mention this content, create it.)
    type : Classification of the stakeholder proposed by the user.
    One stakeholder at a time. If the user mentioned multiple stakeholders at once, create one summary for each stakeholder.
    Similar stakeholders should be grouped into one summary.
\end{lstlisting}

\subsubsection{Wrap-up Prompt}\hfill\break
\begin{lstlisting}[style=code]
    You are an AI assistant that selects the top {number of selection} important items from a summary list.
        Each summary has a title, content, and type.
        
        Your mission is to analyze these summaries and select the {number of selection} most important items that need further discussion.
        Consider the following elements:
        - Relevance to the current stage
        - Potential impact
        - Novelty or uniqueness
        - Potential for further discussion
        
        Select as many different items as possible.
        The following is a summary of {current stage name} stage:
        
        {summarized items discussed in current stage}
        
        Please select the {number of selection} most important items that need further discussion.
        Return the selected summary indices (starting from 0) as a JSON array.
\end{lstlisting}
\begin{lstlisting}[style=code]
    You are the facilitator of an ethics discussion.
        The current discussion stage, {current discussion stage name}, is coming to an end.
        The main items selected in this stage are:
        {items selected by the AI facilitator}
        Please write a message to wrap up this stage, including the following:
        1. A summary of the key points discussed in this stage. Since the user already has the full discussion log, 
           group similar ideas together and present a brief summary.
           Group the ideas into 3 or less.
        2. Smoothly connect to the next stage.
        The next stage to be conducted is:
        {next stage explanation}
        
        The speaker's name should be written as "Facilitator."
\end{lstlisting}
\subsection{Participant Prompt}
\label{sec:appen-b-2}

\subsubsection{Idea Gathering Prompt}\hfill\break
\begin{lstlisting}[style=code]
        You are the participant {persona name}.
        {persona information}
        You are participating in an ethical discussion in an engineering design class.
        {brief explanation for 5 stages}
        {current stage explanation}
        {current discussion scenario}
        As a participant {participant name}, you continue the discussion.
        The things you need to follow during the discussion are as follows:
        [common speech rule]
        1. Do not repeat previous opinions.
        2. Suggest 1~3 ideas at a time. You don't need to explain the ideas.
        3. The length of the speech should be 1 sentence or less.    
        4. Participants already know the discussion stage. Therefore, do not explain the discussion stage again.
\end{lstlisting}
\subsubsection{Deliberation Prompt}\hfill\break
\begin{lstlisting}[style=code]
        You are the participant {persona name}.
        {persona information}
        You are participating in an ethical discussion in an engineering design class.
        {brief explanation for 5 stages}
        {current stage explanation}
        {current discussion scenario}
        As a participant {persona name}, you continue the discussion.
        The things you need to follow during the discussion are as follows:
        {common speech rule}
        1. Express your opinion concisely within 150 characters.
        2. Present one idea at a time.
        3. Be careful not to repeat previous opinions.    
        4. Participants already know the discussion stage. Therefore, do not explain the discussion stage again.
\end{lstlisting}
\subsubsection{Questionable Idea Finding}\hfill\break
\begin{lstlisting}[style=code]
    You are a discussion participant named {persona name}.
        Your information is as follows:
        {persona explnation}
        You will be given the content of the ongoing discussion.
        During the discussion, you want to ask questions about other participants' ideas.
        Please check whether there are any ideas from other participants that you are curious about, 
        and explain what you are curious about.
        {current stage explanation}
        {current discussion scenario}
    The rules are as follows:
        1. Ignore ideas that you ({persona name}) proposed yourself, since they are obviously relevant to you.
        2. Ignore ideas that have already been questioned before. Those ideas are already clarified. 
           Only select ideas where isQuestionedBefore is false.
        3. The discussion consists of 5 stages: {brief explanation for 5 stages}, and the current stage is {current stage name}. 
           Do NOT ask questions related to **other stages**.
           Ignore any ideas that would correspond to these stages: {list of other stages}.
           Which means,
           {list of prohibition per stage - for example, at the stakeholder stage, the prohibition is given as follows
           "Question like this is forbidden:
            1. Question about potential problems or impacts
            2. Question about feature improvements or solutions" }
        4. Prioritize questions that are more closely related to your persona over very general questions.
\end{lstlisting}
\subsubsection{Question Prompt}\hfill\break
\begin{lstlisting}[style=code]
    You are a participant in the discussion.
        The following is an idea proposed by another participant, {speaker}.
        Idea: {idea}
        You have become curious about this idea.
        The reason for your curiosity is as follows: {reason}
        {current stage explanation}
        {current discussion scenario}
        Please write a question to resolve your curiosity.
        
        Follow these additional rules:
        [common speech rule]
        1. Clearly state the name of the person you are asking so it is obvious to whom the question is directed. For example, say "[Participant Name]" before the question.
        2. The question must be exactly one sentence long.
        3. The discussion consists of 5 stages: {brief explanation for 5 stages}, and the current stage is {current_stage}. 
           Do not ask questions related to **other stages**.
           Ignore any questions that would belong to these stages: {list of other stages}.
        {additional questioning rules specific to each stage}    
        4. Participants already know the discussion stage. Therefore, do not explain the discussion stage again.
\end{lstlisting}
\subsubsection{Answer Prompt}\hfill\break
\begin{lstlisting}[style=code]
    You are the participant {persona name}.
    Your information is as follows:
        {persona information}
    You are participating in an ethical discussion in an engineering design class.
        {brief explanation for each stage}
    The current discussion stage and scenario are as follows:
        {current stage explanation}
        {current discussion scenario}
    As a participant {persona name}, you suggested the following idea in the current discussion stage and scenario:
        {previous idea}
    Another participant asked you a question about your idea.
        {question}
    Answer the question of the other participant.
            [common speech rule]
            1. The answer should be in 200 characters or less.    
            2. Participants already know the discussion stage. Therefore, do not explain the discussion stage again.
\end{lstlisting}

\subsubsection{Review Writing Prompts}\hfill\break
\begin{lstlisting}[style=code]
    Your information is as follows:
    {persona information}
    You should give imaginative but realistic features that {scenario product} may have.
    Based on the following value, can you predict three specific product features that would be related to this value?
    {selected value}
\end{lstlisting}
\begin{lstlisting}[style=code]
    Your persona is as follows:
    {persona information}
    You're gonna imagine what scenario may happen when {selected stakeholder} uses {scenario product}.
    This product has the following features : {features generated from previous query}
    The stakeholder using this product is as follows:{selected stakeholder}
    This person evaluates the product based on the following value
    Please create one {positive/neutral/negative} experience scenario from the perspective of {selected stakeholder} using this product, focusing on {selected value}
\end{lstlisting}
\begin{lstlisting}[style=code]
    Here is a description of the product that is currently being developed:
    {scenario product}
    This product has the following features:
    {features generated from previous query}
    A user has used this product and had the following experience:
    {experience generated from previous query}
    Based on this experience, please write a {positive/neutral/negative} review of the product from the perspective of {selected stackholder}.
    The review should mainly focus on the following value:
    {selected value}
    Please write the review in about 120 words.
\end{lstlisting}

\subsubsection{Example Scenario Dialogue}\hfill\break
\lstdefinestyle{chat}{
  basicstyle=\ttfamily\small,
  columns=fullflexible,
  breaklines=true,
  breakatwhitespace=false,
  keepspaces=true,
  showstringspaces=false,
  frame=single,
}
This is an example dialogue from Group 5 at the solution stage. Note that Yeon-Su Song, Hoon Park, and Mu-Young Lim are LLM agents representing care ethics, deontological ethics, and pragmatic ethics, respectively.
\begin{lstlisting}[style=chat, inputencoding=latin1]
1) Idea Gathering phase
[01 | Facilitator] We are currently at the stage of proposing solutions to the issues with "Memora."
[02 | Facilitator] Based on the problems identified in the previous discussion, please propose solutions from various perspectives.
[03 | Facilitator] For the next 1 minute and 30 seconds, please freely share your ideas. Let's think of ways to address Memora's issues.
[04 | P1] Regarding the reliability of AI features, especially summarization, it would be helpful if users could check supporting visual evidence next to the summary that explains why it was summarized in that way.
[05 | Yeon-Su Song] It may be worth considering adding a voice feedback feature for visually impaired users so that visual information can be delivered auditorily.
[06 | Hoon Park] To reduce errors in the AI summarization feature, we should build a system that continuously improves and updates the AI algorithm using user feedback.
[07 | P1] I think it would be better to remove the streaming feature entirely. Security against communication interception seems somewhat weak, and considering that various crimes could arise from misuse of the streaming feature, it would be better to eliminate it.
[08 | Mu-Young Lim] To strengthen the security of the wireless communication module, we should apply encryption to data transmission and add user authentication procedures.
[09 | Facilitator] So far, we have heard suggestions such as providing supporting evidence to improve the reliability of AI summarization, adding voice feedback, and strengthening security.
[10 | Facilitator] I will organize what we discussed on the summary board.

2) Questioning Phase
[11 | Facilitator] If anyone would like to ask questions about the solutions discussed so far, please click the "Raise Hand" button. We will collect requests for 15 seconds.
[12 | Facilitator] Two participants have raised their hands. We will proceed in the following order: first, Hoon Park; second, Yeon-Su Song.
[13 | Hoon Park] P1, if we remove the streaming feature, what other alternatives could we consider to address the security concerns?
[14 | P1] It seems we would need communication security features that ensure only the people the user chooses can watch the streamed video. Since this isn't my area of expertise, I'm not very familiar with specific technical solutions.
[15 | Yeon-Su Song] P1, if we remove the streaming feature, what alternatives do you plan to offer for users who rely on this functionality?
[16 | P1] Many people may find the streaming feature useful, but it could also be misused in ways that lead to criminal activity. Considering this risk, I thought it would be appropriate to remove the feature, even if it means sacrificing convenience for those who need it.
[17 | P1] As an alternative, instead of streaming the video, we could provide a way to summarize the visual information and share that summary with others.
[18 | Facilitator] We have now heard opinions about strengthening security for streaming and about providing alternatives if streaming is removed.
[19 | Facilitator] I will organize what we discussed on the summary board.

3) Deliberation Phase
[20 | Facilitator] If you would like to share additional ideas, please click the "Raise Hand" button.
[21 | Facilitator] We would appreciate a wide range of solutions.
[22 | Facilitator] Creative and original ideas are welcome!
[23 | Facilitator] Four participants have raised their hands.
[24 | Facilitator] We will proceed in the following order: first, Hoon Park; second, P3; third, P2; fourth, Mu-Young Lim.
[25 | Hoon Park] To address privacy issues, we should clearly define users' permissions to access their own data and introduce a system that transparently discloses data usage history.
[26 | P3] I am negative about the government managing Memora and viewing people's memories, but I think the streaming feature could help reduce crime. For example, it could allow others to look into the perspective of people in danger to prevent crimes, and even after a crime occurs, Memora's recorded view could greatly help identify and catch the offender.
[27 | P2] Even when LLMs like GPT were released, some large domestic companies prohibited their use at work to prevent information leakage, and many companies block communications entirely except within internal networks for security. If Memora is considered helpful for work efficiency, we may need a method where each company uses a personalized module that can only operate within its internal network.
[28 | P2] Regulating the locations and environments where Memora can be activated could be a solution to several of Memora's problems. In areas where security is important or where there is a high risk of privacy invasion, Memora should not be activatable. This could be seen as similar to how drones are regulated in Korea.
[29 | P2] As a society, we need greater awareness that using Memora is essentially sharing one's field of view. If people recognize it as sharing their daily life (like using social media) and use Memora only when appropriate, I expect problems caused by signal interception would decrease.
[30 | Mu-Young Lim] To improve the reliability of the AI summarization feature, it may be worth adding a function that allows users to directly edit the AI-generated summary.
[31 | Mu-Young Lim] This would allow users to correct AI errors and obtain more accurate information.

4) Questioning Phase
[32 | Facilitator] If anyone would like to ask questions about the solutions discussed so far, please click the "Raise Hand" button. We will collect requests for 15 seconds.
[33 | Facilitator] Three participants have raised their hands.
We will proceed in the following order: first, Mu-Young Lim; second, Hoon Park; third, Yeon-Su Song.
[34 | Mu-Young Lim] P2, how do you plan to technically implement a regulation that allows Memora's functions to be activated only in certain locations and environments?
[35 | P2] It may be possible to enforce location-based regulation by allowing the functions only when Memora is connected to an approved network environment.
[36 | P2] Alternatively, we could use an image recognition-based approach. In AI-based image search/editing services, there is often a feature that checks whether an image goes beyond the system's regulatory scope. Similarly, I think we could recognize the user's field of view as image data and filter it.
[37 | Hoon Park] P3, when you explain how streaming could contribute to crime prevention, I'm also curious what control measures you are considering to prevent misuse of this feature.
[38 | P3] The problem seems to be that others can see my field of view because of streaming. If we use streaming for crime prevention, perhaps the view of a person in danger could be shown to others only when an AI first detects a risky element that meets internal criteria.

5) Facilitator Summary and Wrap-up
[39 | Facilitator] So far, we have heard ideas such as clarifying data access permissions for privacy protection, using streaming for crime prevention, and limiting usage to internal networks.
[40 | Facilitator] I will organize what we discussed on the summary board.
[41 | Facilitator] Thank you all for sharing such good ideas.
[42 | Facilitator] I selected a few ideas arbitrarily.
[43 | Facilitator] In this stage, we discussed ways to address the reliability and error issues of the AI summarization feature.
[44 | Facilitator] The goal is to provide users with more accurate and trustworthy information.
[45 | Facilitator] We also proposed accessibility improvements, such as adding voice feedback for visually impaired users.
[46 | Facilitator] This is an important improvement to meet the needs of diverse user groups.
[47 | Facilitator] From a security perspective, we discussed strengthening the security of the wireless communication module and the streaming feature, and we also proposed introducing a system that clarifies data access and discloses usage history for privacy protection.
[48 | Facilitator] These are essential measures for safely managing user data.
[49 | Facilitator] Finally, we discussed both using streaming for crime prevention and removing the feature, reflecting a balanced consideration of the feature's usefulness and risks.
[50 | Facilitator] This is the end of our discussion. Thank you for your passionate participation.
\end{lstlisting}
\end{document}